# Quantum decoherence dynamics of divacancy spins in silicon carbide

Hosung Seo[1], Abram L. Falk[1,2], Paul V. Klimov[1], Kevin C. Miao[1], Giulia Galli[1,3],

and David D. Awschalom*[1]

1. The Institute for Molecular Engineering, The University of Chicago, Chicago, IL 60615, USA
2. IBM T. J. Watson Research Center, Yorktown Heights, NY 10598, USA
3. Materials Science Division, Argonne National Laboratory, Lemont, IL 60439, USA

## Abstract

Long coherence times are key to the performance of quantum bits (qubits). Here, we experimentally and theoretically show that the Hahn-echo coherence time of electron spins associated with divacancy defects in 4$H$-SiC reaches 1.3 ms, one of the longest Hahn-echo coherence times of an electron spin in a naturally isotopic crystal. Using a first-principles microscopic quantum-bath model, we find that two factors determine the unusually robust coherence. First, in the presence of moderate magnetic fields (30 mT and above), the $^{29}$Si and $^{13}$C paramagnetic nuclear spin baths are decoupled. In addition, because SiC is a binary crystal, homo-nuclear spin pairs are both diluted and forbidden from forming strongly coupled, nearest-neighbor spin pairs. Longer neighbor distances result in fewer nuclear spin flip-flops, a less fluctuating intra-crystalline magnetic environment, and thus a longer coherence time. Our results point to polyatomic crystals as promising hosts for coherent qubits in the solid state.

## Introduction

Impurity-based electron spins in crystals, such as the nitrogen vacancy (NV) center in diamond[1,2], donor spins in silicon[3], transition-metal ions[4], and rare-earth ions[5] have recently attracted great interest as versatile solid-state qubits. Among the key measures for qubit performance, coherence times characterize the lifetime of a qubit. In quantum computing, long spin coherence times are necessary for executing quantum algorithms with many gates[6]. Qubits with robust coherence are also ideal systems for developing applications such as collective quantum memories[7] and nano-scale quantum sensors[8,9]. Nonetheless, interactions between the spin qubit and the bath of paramagnetic nuclei in the crystal eventually limit the qubit's coherence[10-12]. One of the standard measures of spin coherence time is the ensemble Hahn-echo coherence time ($T_2$)[13]. For NV centers in naturally isotopic diamond and for donor spins in natural silicon, $T_2$ times have been measured to be 0.63 ms[14] and 0.5 to 0.8 ms[15-17], respectively. These are set by the presence of naturally occurring $^{13}$C (1.1%, $I_C$=1/2) isotopes[11,12,18-22] and $^{29}$Si (4.7%, $I_{Si}$=1/2) isotopes[10,23-25]. For Mn:ZnO, a 0.8-ms $T_2$ time has been reported[4], which is set by the $^{67}$Zn (4.1%, $I_{Zn}$=5/2) isotopic concentration.

Several techniques can be used to extend spin coherence, including isotopic purification[12,25], dynamical decoupling[26-28], and the use of particular 'clock transitions' that are immune to external magnetic perturbations[29-31]. These techniques cannot be used in all applications, however, and moreover, the extent to which spin coherence can be extended is typically correlated to the original $T_2$ time. Therefore, the Hahn-echo $T_2$ time in a naturally isotopic crystal remains an important metric for qubit performance.

Recently, Christle *et al.* reported a $T_2$ time of 1.2 ms for divacancies in SiC[32], which are spin-1 defects[33-42]. However, the spin dynamics underlying this coherence time were not understood. Naturally isotopic SiC contains both $^{29}$Si (4.7%) and $^{13}$C (1.1%) isotopes. Nevertheless, in spite of having a higher nuclear spin density than natural diamond, SiC was able to host qubits with a much longer $T_2$ time than those of NV centers, implying a suppression of nuclear spin bath fluctuations. Yang *et al.* recently published an insightful theoretical paper[43] on the nuclear-bath driven decoherence of single silicon vacancy($V_{Si}$) in SiC, a spin-3/2 defect[44-50]. Using the cluster-correlation expansion (CCE) theory[51], they showed that heterogeneous nuclear spin flip-flop processes are suppressed in SiC due to the difference between the gyromagnetic ratios of $^{29}$Si and $^{13}$C nuclear spins (or heterogeneity). Similar heterogeneity and bath decoupling effects were also discussed for GaAs quantum dots[52]. Based on the bath decoupling effect, Yang *et al.*, suggested that the spin coherence time in naturally isotopic SiC would be longer than that of the NV center in diamond[43]. However, direct experimental verification in SiC has been challenging using single $V_{Si}$ spins[48,53], partly because hyperfine coupling to the $S = 3/2$ state gives rise to irregular coherence patterns[43].

Here, we combine experiment and theory to study the decoherence dynamics of the $S$=1 electronic spin ensemble of the neutral ($kk$)-divacancy in 4$H$-SiC over a wide range of magnetic fields. We use optically detected magnetic resonance (ODMR)[36] and a first-principles microscopic quantum-bath model[54] combined with the CCE method[51,52] to demonstrate that the $T_2$ time of the divacancy spin in 4$H$-SiC can reach 1.3 ms, an unusually long $T_2$ time. Our theoretical results successfully explain all the important features found in our experiment such as the behavior of $T_2$ as a function of magnetic field and the fine details in the electron spin echo envelop modulations (ESEEM)[13]. In particular, by studying ensembles of $S = 1$ centers instead of single $S = 3/2$ centers, we provide strong evidence that in SiC, the Si and C nuclear spin baths are decoupled at moderate magnetic field (~30 mT), confirming the predictions of Yang et al.[43]. In addition to verifying Yang's predictions, we show that a key factor underlying the long coherence times in SiC is the fact that homo-nuclear spin pairs in this binary crystal must be at least two lattice sites away from each other. This separation limits the strength, and therefore the flip-flop rate, of the most strongly coupled spin pairs.

## Results

**Optically detected spin coherence in SiC.**

Our experiments use 4*H*-SiC wafers (purchased from Cree, Inc.) with vacancy complexes intentionally incorporated during crystal growth. The divacancy density is approximately $10^{12}$ cm$^{-3}$ [37]. In this study, we consider the (*kk*)-divacancy[36,37], which is schematically shown in Fig. 1. We use a 975 nm laser diode to illuminate the sample, which, through ODMR, polarizes the electronic ground state of the divacancies into their $m_s = 0$ state[36,37]. The divacancies exhibit more intense photoluminescence (PL) in their $m_s = \pm 1$ state[36,37] than in their $m_s = 0$ state, allowing the spin of the defects to be read out via the PL intensity. We use a movable permanent magnet to apply a *c*-axis oriented magnetic field (*B*)[36]. To measure the pure spin dephasing rate, we perform standard Hahn-echo pulse sequence [$\pi/2$ pulse – $t_{\text{free}}/2$ – $\pi$ pulse – $t_{\text{free}}/2$ – $\pi/2$ pulse][13] measurements. The first $\pi/2$ pulse creates a superposition of the $m_s = +1$ and $m_s = 0$ states, and the following $\pi$ pulse reverts the spin precession after the $t_{\text{free}}/2$ free evolution. At the end of the Hahn echo sequence, the spin coherence is refocused, removing the effects of static magnetic inhomogeneity. The last $\pi/2$ pulse converts the phase difference in the superposition state to a population difference in the $m_s = +1$ and $m_s = 0$ states, which we then measure through a change in the PL intensity.

In Fig. 2, we show the measured Hahn-echo coherence of the divacancy ensemble at three representative magnetic fields and as a continuous function of magnetic field. At low magnetic fields, e.g. 2.5 mT and 6.5 mT shown in Fig. 2 (a), the spin coherence rapidly collapses and revives as a function of time. Simultaneously, its envelop decays over time, leading to the loss of coherent phase information within 1 ms. In Fig. 2, we observe that this spin decoherence is largely suppressed and that the coherence is further extended as the static magnetic field is increased. We show the $T_2$ as a function of magnetic field in Fig. 3 (a). We find that $T_2$ increases as a function of magnetic field and saturates to 1.3 ms at a magnetic field of roughly 30 mT. There is a dip in $T_2$ at a magnetic field of ~ 47 mT, which is also visible in Fig. 2 (c) as a coherence drop. This magnetic field converts to 1.31 GHz energy splitting, corresponding to the zero-field splitting of the (*kk*)-divacancy[37]. The coherence drops at this ground state level anti-crossing (GSLAC) as the $m_s = 0$ spin state can significantly mixes with $m_s = -1$ spin sublevel.

**Quantum bath approach to decoherence.**

To understand the decoherence dynamics observed in experiment, we use quantum bath theory, which describes the qubit decoherence occurring due to the entanglement between the qubit and the environment[54]. We apply the same theory to the NV center and to the (*kk*)-divacancy spin so as to compare results consistently and to understand the underlying physical reasons responsible for their difference. The two defects share many common features[34-36,39]. For example, the *c*-axis oriented (*kk*)-divacancy (Fig. 1 (a)) exhibits the same $C_{3v}$ point-group symmetry and $^3A_2$ spin triplet ground state as the NV center in diamond (Fig. 1 (b)). Furthermore, similar to the NV center, the divacancy ground state is mainly derived from the three carbon $sp^3$ orbitals localized around the silicon vacancy site in SiC. The only difference between the divacancy-in-SiC model and the NV-center-in-diamond model is the type of nuclear spin bath along with their lattice structures as shown in Fig. 1 (a) and (b), respectively. We note that the dynamics of NV-center decoherence has been well-understood, and that our results are in excellent agreement with those previously reported in the literature[18,19,22]. In our model, we ignore any possible effects arising from the nuclear and electronic spin-lattice relaxation. (See Supplementary Note 1 for further discussions). To solve the central spin model, we use the CCE method[51,52], and we systematically approximate the coherence function at different orders. No adjustable parameters are used. Further details on the theoretical methods and the numerical calculations can be found in the methods section and the Supplementary Notes 1-3, together with Supplementary Figures 1-8 and Supplementary Table 1.

In Fig. 2 (b) and (d), we show the theoretical Hahn-echo coherence functions of the divacancy spin, to be compared with the experimental coherence data shown in Fig. 2 (a) and (c), respectively: the agreement between theory and experiment is excellent. In Fig. 3 (a), we compare the theoretical $T_2$ times of the divacancy to the experimentally measured $T_2$ times. Both $T_2$ curves rapidly increase as a function of the free evolution time ($t_{\text{free}}$) up to a magnetic field of 20 mT. For $B > 30$ mT, they both saturate at a limit of 1.3 ms, although the experimental $T_2$ curve appears to saturate more slowly. The dip in $T_2$ at a magnetic field around 47 mT is not found in the theory, because in our model, we did not consider spin mixing between $m_s = 0$ and $m_s = -1$ near the GSLAC. As a verification of our methods, we also compare the computed and measured divacancy $T_2$ times with the theoretical $T_2$ times of the NV center in diamond (Fig. 3 (a)). The theoretical limit of the NV-center $T_2$ time is found to be about 0.86 ms, in agreement with ensembles measurements[14] and with previous theoretical results obtained by the disjoint-cluster method[18] and an analytical method[22]. Our theoretical results confirm that the divacancy $T_2$ time in naturally isotopic 4*H*-SiC is much longer than that of the NV center in naturally isotopic diamond.

In Fig. 3 (b), we compare the theoretical and experimental coherence functions at two different magnetic fields (12.5 mT and 17.5 mT). We find that the measured oscillation pattern of the coherence is also well reproduced by the theory, including the relative peak height and width, further verifying our microscopic model comprising $^{29}$Si and $^{13}$C nuclear spins. In the presence of a static magnetic field, the $^{29}$Si and $^{13}$C nuclear spins precess at their respective Larmor frequencies and induce electron spin echo envelop modulation (ESEEM)[13,55]. In Fig. 3 (c) and (d), we compare the *B*-normalized fast Fourier transform (FFT) spectra of the full experimental and theoretical coherence functions shown in Fig. 2 (c) and (d), respectively. Two-peak structures are clearly seen, centered at the $^{29}$Si and $^{13}$C nuclear gyromagnetic ratios, which are 8.7 MHz/T and 10.9 MHz/T in experiment, and 8.5 MHz/T and 10.7 MHz/T in theory, respectively. In addition to the Larmor-frequency peaks, we observe faint, but appreciable hyperbolic features both in experiment and theory as denoted by dotted arrows in Fig. 3 (c) and (d), respectively.

Since the ESEEM spectrum is derived from the independent precession of nuclear spins, the generic features of the spectrum may be understood using the analytical solution of an independent nuclear spin model (see Supplementary Fig. 5)[13,55]:

$$\mathcal{L}_{\text{ESEEM}}(t_{\text{free}}) = \prod_i (1 - 2k_i \sin^2(w_i t_{\text{free}}/4) \sin^2(a_i t_{\text{free}}/4)), \qquad (1)$$

where *i* labels individual $^{29}$Si and $^{13}$C nuclear spins in the nuclear spin bath, $k_i$ is a modulation depth parameter, $w_i$ is the frequency of the $i^{\text{th}}$ nuclear spin, and $a_i$ is a frequency that depends on the hyperfine coupling parameters and the nuclear frequency (see Supplementary Note 3). When the electron spin is in the $m_s$=0 state, the hyperfine field on the nuclear spins is zero, leading to coherence oscillations at the bare nuclear frequencies. For the electron spin in the $m_s$=+1 state, each nuclear spin experiences a different hyperfine field depending on its position relative to the electron spin, giving rise to the hyperfine-frequency term ($a_i$) in Eq. (1). We note that these $a_i$ terms in Eq. (1) due to weak hyperfine interactions give rise to the hyperbolic features found in the FFT spectra shown in Fig 3. (c) and (d). We find similar hyperbolic features in the computed FFT spectrum of the NV center in diamond (not shown), although less pronounced compared to that of the SiC divacancy FFT spectrum. The modulation depth parameter, $k_i$ in Eq. (1) is inversely proportional to the magnetic field (see Supplementary Note 3), explaining the suppression of the oscillation amplitude at a large magnetic field found both in experiment and theory, as shown in Fig. 2 (a) and (b), respectively. The FFT intensities also diminish as *B* is increased for the same reason as shown in Fig. 3 (c) and (d).

**Suppressed qubit decoherence in silicon carbide.**

We now turn our attention to the microscopic origin of the longer $T_2$ time of the divacancy (1.3 ms at $B$ = 30 mT) compared to that of the NV center (0.8 ms at $B$ = 30 mT), in spite of the much larger number of nuclear spins in the SiC lattice. By comparing calculations performed at different CCE orders (see Supplementary Fig. 3), we find that for both NV and the divacancy the computed Hahn-echo coherence time is numerically converged at the CCE-2 level of theory. This finding indicates that the dominant contribution to decoherence comes from pairwise nuclear transitions induced by nuclear dipole-dipole couplings. The decoherence of the NV center in diamond is mainly caused by pair-wise nuclear spin flip-flop transitions (↑↓ ↔ ↓↑), which induce magnetic noise at the NV center through the hyperfine interaction. Other pairwise nuclear spin transitions, such as co-flips (↑↑ ↔ ↓↓), are suppressed at magnetic fields larger than roughly 10 mT. These results agree well with those previously reported for NV centers in diamond[18,19,22].

In 4$H$-SiC, the nuclear spin interactions can be grouped in two categories: heterogeneous, between $^{13}$C and $^{29}$Si, and homogeneous interactions between nuclear spins of the same kind. The Hahn-echo coherence function of the divacancy can then be written as:

$$\mathcal{L}_{(kk)}(t_{\text{free}}) \approx \prod_i \tilde{\mathcal{L}}_i \prod_{\{i,j\}} \tilde{\mathcal{L}}_{i,j} = \prod_i \tilde{\mathcal{L}}_i \prod_{\{i,j\}_{\text{hetero}}} \tilde{\mathcal{L}}_{i,j} \prod_{\{i,j\}_{\text{homo}}} \tilde{\mathcal{L}}_{i,j}, \quad (2)$$

where $\tilde{\mathcal{L}}_i$ is a single-correlation term from the $i^{\text{th}}$ nuclear spin and $\tilde{\mathcal{L}}_{i,j}$ is an irreducible pair-correlation contribution from the $i - j$ nuclear spin pair. The product over $\{i,j\}_{\text{hetero}}$ include all $^{13}$C - $^{29}$Si nuclear spin interactions, while the product over $\{i,j\}_{\text{homo}}$ include all $^{13}$C - $^{13}$C and $^{29}$Si - $^{29}$Si spin pairs. We define the following heterogeneous and homogeneous coherence functions:

$$\mathcal{L}_{\text{hetero}}(t_{\text{free}}) = \prod_i \tilde{\mathcal{L}}_i \prod_{\{i,j\}_{\text{hetero}}} \tilde{\mathcal{L}}_{i,j}, \quad (3)$$

$$\mathcal{L}_{\text{homo}}(t_{\text{free}}) = \prod_i \tilde{\mathcal{L}}_i \prod_{\{i,j\}_{\text{homo}}} \tilde{\mathcal{L}}_{i,j}. \quad (4)$$

To investigate the effect of the heterogeneity, we vary the gyromagnetic ratio of $^{29}$Si ($\gamma_{\text{Si}}$) as a theoretical parameter while that of $^{13}$C ($\gamma_{\text{C}}$) is fixed at the experimental value. In Fig. 4, $\mathcal{L}_{\text{hetero}}$ is shown at four different $\gamma_{\text{Si}}$ values at a magnetic field of 30 mT. We find that there would be a significant decay of $\mathcal{L}_{\text{hetero}}$ if the $^{29}$Si and $^{13}$C gyromagnetic ratios were hypothetically the same ($\Delta_\gamma \equiv \gamma_{\text{C}} - \gamma_{\text{Si}} = 0$), while small differences in the gyromagnetic ratios ($\Delta_\gamma = 0.03$ MHz/T and 0.16 MHz/T for the two middle plots in Fig. 4(a)) are sufficient to significantly suppress the decay. Furthermore, when using the experimental values of $\gamma_{\text{Si}}$ and $\gamma_{\text{C}}$, $\mathcal{L}_{\text{hetero}}$ does not show any envelop decay, indicating no contribution from pair-wise heterogeneous nuclear spin transitions for $B$ > 10 mT. Due to the sign difference between the gyromagnetic ratios of $^{29}$Si and $^{13}$C ($\gamma_{Si} < 0$, $\gamma_C > 0$), when $B > 10$ mT, the lowest-energy $^{29}$Si - $^{13}$C pairwise spin transition is the co-flip of the nuclear spins ($\uparrow\uparrow \leftrightarrow \downarrow\downarrow$). In addition to the hyperfine field difference on the order of few kHz, the difference between $\gamma_{Si}$ and $\gamma_C$ gives an extra Zeeman contribution to the energy gap (~0.2 MHz at $B = 10$ mT) for the co-flips, which is larger than the typical heterogeneous dipole-dipole transition rate (~ kHz) in 4$H$-SiC.

The absence of heterogeneous nuclear spin transitions amounts to a decoupling of the nuclear spin bath in SiC and therefore the Hahn-echo coherence function is given by:

$$\mathcal{L}_{(kk)}(t_{\text{free}}) \approx \mathcal{L}_{\text{homo}} = \mathcal{L}_{^{29}\text{Si}} \mathcal{L}_{^{13}\text{C}}, \tag{5}$$

where $\mathcal{L}_{^{29}\text{Si}}$ and $\mathcal{L}_{^{13}\text{C}}$ are the Hahn-echo coherence functions of the divacancy spin coupled to $^{29}$Si nuclear spins only and to $^{13}$C nuclear spins only, respectively. Since only transitions between homonuclear spins contribute to $\mathcal{L}_{(kk)}$, the density of nuclear spins contributing to the electron spin decoherence turns out to be similar to that found in diamond[53], in spite of the total density of spins being much higher. However, this so-called dilution effect by itself would point to a similar electron spin decoherence rate in SiC and in diamond[53], contrary to what is found experimentally (1.3-ms and 0.63-ms $T_2$ time in SiC and diamond, respectively).

To better understand the nature of the nuclear spin baths in SiC, we compare in Fig. 4 (b) the ensemble-averaged numbers of homogeneous nuclear spin pairs that are contributing to the decoherence of the divacancy in 4$H$-SiC and of the NV center in diamond. In the former case, the homogeneous $^{29}$Si (4.7%) spin pairs are the dominant source of the qubit decoherence, and their number is larger than that of the $^{13}$C (1.1%) spin pairs in diamond. However, being further apart, their contribution is weaker than that of the homonuclear spin pairs in diamond. In Fig. 4 (c) the distributions of nuclear spin pairs shown in Fig. 4 (b), are reported as a function of nuclear-nuclear distance. In the case of the NV center in diamond, there is a small but significant number of nuclear spin pairs at a distance less than 3.0 Å, including first-, second-, and third nearest C-C neighbors. These spins exhibit strong secular dipole-dipole transition rates, ranging from 0.24 kHz to 2.06 kHz: while they are minority spin pairs in number, they account for more than 90% of the coherence decay for the NV center in diamond (see Supplementary Fig. 2 (e)). In contrast, in 4$H$-SiC, the smallest distance between homogeneous spins is 3.1 Å, corresponding to the Si-Si or C-C neighbors in SiC. As a result, the secular dipole-dipole transition rates for all the homogeneous nuclear spin pairs in 4$H$-SiC turn out to be less than 0.08 kHz. Our results show that the absence of strongly coupled nuclear spin clusters in SiC plays a key role in explaining the surprisingly long divacancy $T_2$ times.

**Isotopic purification to lengthen $T_2$.**

We showed that the coherence time of the divacancy in our naturally isotopic, semi-insulating 4$H$-SiC is 1.3 ms. In principle, the $^{29}$Si or $^{13}$C nuclei can be removed by isotopic purification, which is available in SiC[56,57], and a longer qubit coherence time could be achieved[12,18,24,58]. In Fig. 5, we report the Hahn-echo $T_2$ of the divacancy ensemble in 4$H$-SiC computed as a function of the $^{13}$C concentration, while that of $^{29}$Si was fixed at given values, and we compare the results with those for the Hahn-echo $T_2$ of the NV center in diamond. In the case of the NV center (Fig. 5 (f)), we find that $T_2$ scales as $1/n_c$ ($T_2 \approx 0.95(n_C)^{-1.08}$), where $n_c$ is the concentration of the $^{13}$C isotopes, in excellent agreement with previous theoretical[18] and experimental[11] findings.

In 4$H$-SiC, we observe that the divacancy $T_2$ time increases as both $^{29}$Si and $^{13}$C concentrations are reduced. However, this increase does not appear to follow a simple power-law scaling behavior. For example, in Fig. 5 (a), where the $^{29}$Si concentration is fixed at the experimental value of 4.7%, $T_2$ is nearly constant as the $^{13}$C concentration is lowered below 1.1%. The behavior of $T_2$ is also significantly dependent on the applied magnetic field. We note that even if the $^{13}$C concentration is reduced, $^{29}$Si nuclear spins are still the majority ones, and thus responsible for limiting the coherence time. As the $^{29}$Si concentration is reduced from 4.7% to 0% (Fig. 5 (a) to Fig. 5 (e)), the behavior of $T_2$ as a function of $^{13}$C concentration becomes linear, similar to that of the NV center in diamond. To rationalize the scaling behavior of the divacancy $T_2$, we compute the dependence of $\mathcal{L}_{^{13}C}$ and $\mathcal{L}_{^{29}Si}$ on the $^{13}$C and $^{29}$Si concentrations using Eq. (5), respectively, which we then fit with the compressed exponential decay function, ($e^{-\left(\frac{t_{free}}{T_2}\right)^n}$). We find that $T_2$ time of $\mathcal{L}_{^{29}Si}$ and $\mathcal{L}_{^{13}C}$ follows a simple scaling law as a function of nuclear spin concentration: $T_{2,Si} \approx a_{Si}(n_{Si})^{N_{Si}}$ and $T_{2,C} \approx a_C(n_C)^{N_C}$, with $a_{Si}$ = 4.27 ms, $N_{Si}$ = -0.74, $a_C$ = 3.31 ms, and $N_C$ = -0.86, and the stretching exponent ($n$) is ~ 2.6 for both C and Si when $B$ > 30 mT. This exponent is the same as that of the total coherence function, and although in good agreement with experiments (2.3), it is slightly larger. Using Eq. (5), we thus find that the divacancy $T_2$ scales as follows:

$$T_2 \approx \left[\left(a_{Si} n_{Si}^{N_{Si}}\right)^{-n} + \left(a_C n_C^{N_C}\right)^{-n}\right]^{-1/n}, \qquad (6)$$

Eq. (6), plotted as a dashed line in Fig. 5 (a) to 5 (f), describes very accurately our full numerical simulation results at magnetic fields larger than 20 mT. As noted above, however, the scaling behavior significantly changes as the magnetic field is decreased under 20 mT and it cannot be described by Eq. (6). The inadequacy of Eq. (6) at low magnetic fields stems from the fact that heterogeneous nuclear spin transitions may occur, further limiting the $T_2$ times. Therefore, the decoupling effect leading to Eq. (5) and thus, the scaling law in Eq. (6) are invalid at low magnetic fields.

# DISCUSSION

We used a combined experimental and theoretical study to investigate the decoherence dynamics of divacancy spin qubits in 4$H$-SiC. We showed that, for $B > 30$ mT at $T = 20$ K, the $T_2$ time of the divacancy reaches 1.3 ms, almost two times longer than that of the NV centre. Using a combined microscopic quantum bath model and a CCE computational technique, we found that 1.3 ms corresponds to the theoretical limit imposed by the presence of nuclear spins from naturally occurring $^{29}$Si and $^{13}$C isotopes. This limit is much longer than the corresponding one for the NV center, which is ~ 0.86 ms. The long spin coherence in SiC stems from the combination of two effects: the decoupling of the $^{13}$C and $^{29}$Si spin baths at a finite magnetic field, and the presence of active spins much further apart than those in diamond (for example, the closest ones belong to second neighbors in SiC and to first neighbors in diamond). We showed that, while the coherence of the NV center is mainly limited by a few strongly interacting nuclear spin pairs belonging to nuclei within ~ 3.0 Å of each other, in SiC, the homo-nuclear spin pair interactions are much weaker as they belong to second or further neighbors (see Fig. 1 (a)). We note that the absence of strongly interacting nuclear spins in SiC is not a simple dilution effect. For example, the nuclear spin density in natural diamond is very low (1.1%), i.e. it can be considered a diluted bath. Nevertheless, the distance between nuclei is such that strong nuclear spin interactions may arise, contributing to the decoherence of the NV center in diamond. In SiC, Si and C spins have a much larger minimal distance from each other.

All experiments were performed at a low temperature (T = 20 K) to exclude thermal effects and to focus on the pure dephasing of the divacancy spin (see Supplementary Note 1 for further discussions). Upon an increase of temperature, however, the divacancy $T_2$ time would decrease significantly, as demonstrated in previous work[37]. In Ref. 37, at low field, the $T_2$ time of the divacancy spin was observed to decrease from 360 μs at 20 K to 50 μs at room temperature. In contrast, the NV-center coherence has been known to be relatively insensitive to a temperature change, thus a long coherence time can be measured even at room temperature[14]. The insensitivity of the NV-center coherence to temperature has been mainly attributed to the high Debye temperature and small spin-orbit coupling in diamond. However, the origin of the temperature dependence of the divacancy coherence in SiC is yet unknown.

Although overall, our theoretical and experimental results are in excellent agreement, we did find a few minor discrepancies. First, the ESEEM frequencies in experiment are blue-shifted by about 0.2 MHz/T from the free $^{13}$C and $^{29}$Si frequencies. The blue-shift effect becomes prominent in the appearance of the coherence oscillation at a low magnetic field such as $B = 2.5$ mT in Fig. 2 (a). When compared to the corresponding theoretical plot in Fig. 2b, the ESEEM peaks appear slightly faster in the experiment. Two possible reasons for the blue-shift of the ESEEM frequencies could be the presence of a stray transverse magnetic field[18] and the presence of non-secular Zeeman and hyperfine interactions[21], which our theory does not consider (see Supplementary Note 1 for further details). Second, we found that the stretching exponent, determined from fits of the coherence decay is 2.3 in experiment, and 2.6 in theory. For the NV center, our model yields 1.9, which is in a good agreement with previous analytical calculations[22]. Experimentally, in diamond, decay exponent ranging from 1.2 to 2.7 were reported[14], depending on the sample and the *B*-field misalignment. Finally, the theoretical divacancy $T_2$ times also saturate at a smaller *B* field than the experimental $T_2$ times, for reasons we do not understand.

In this study, we considered the coherence of divacancy spin ensembles. However, the divacancy decoherence dynamics at the single-spin level is also of interest. In Supplementary Fig. 4, we show the variation of the divacancy single-spin $T_2$ time in random nuclear spin environments compared to that of the NV center in diamond. We find that the divacancy single-spin $T_2$ ranges from 0.6 ms to 1.7 ms at a magnetic field of 11.5 mT, while it ranges from 0.4 ms to 1.4 ms at $B$ = 11.5 mT for the NV center in diamond. Similar to the NV center in diamond, the divacancy single-spin coherence dynamics could show a rich complex dynamics depending on individual local nuclear spin environments. Other important factors for the single-spin coherence in SiC may include the effects of strain, thermal, magnetic, and electric inhomogeneities.

Our combined experimental and theoretical work lays a solid foundation to understand the robust divacancy spin coherence. The essential physics should apply to other potential spin qubits in SiC as well, thus providing a benchmark for future implementation of other spin qubits in this material[59-61]. Moreover, our model has implications beyond the crystal studied in this effort. The dynamics responsible for the coherence found in SiC, a binary crystal, may allow qubits in ternary and quaternary crystals to have even longer spin coherence times. For example, our results suggest that alloying the SiC lattice with larger elements such as Ge may further extend the coherence time of the divacancy spins. Since substitutional Ge would replace some $^{29}$Si atoms, it could serve as an alternative path to isotopic purification, especially for applications that require a large number of coherent spins. In addition, interesting host crystals with useful functionalities are normally found in binary or ternary crystals such as carbides, nitrides and oxides[59,62]. The piezoelectricity in AlN is one example. Complex oxides can exhibit exotic collective behaviors such as ferroelectricity, ferromagnetism, and superconducting behavior. Combining these collective degrees of freedom with coherent spin control in complex materials would be a promising route to hybrid quantum systems.

## METHODS
**Experimental methods**

As described in the main text, the 4*H*-SiC samples are high-purity semi-insulating wafers purchased from Cree, Inc (part number: W4TRD0R-0200). Since they contain "off-the-shelf" neutral divacancies, we dice them into chips and measure them without any further sample preparation. The SiC samples are 3-4 mm chips attached to coplanar microwave striplines with rubber cement. In turn, the microwave stripline is soldered to a copper cold finger, which is cooled by a Janis flow cryostat.

For ODMR measurements, we use a 300 mW, 1.27 eV (975 nm) diode laser, purchased from Thorlabs, Inc. 60 mW reaches the sample. We focus the laser excitation onto the sample using a 14 mm lens and collect the photoluminescence (PL) using that same lens. We then focus the collected PL onto an InGaAs photoreceiver, which was purchased from FEMTO, a German electronics manufacturer. Although we did ensemble measurement, it may be worth commenting on the count rates achieved in as-received samples. When single defects were considered in our previous study[32], we observed count rates of 3-5 kcts. However, because we were using a lower efficiency measurement apparatus than the avalanche photodiodes used for diamonds, this should not be directly compared to the 20-30 kcts of a typical NV center. To gate the laser during the Hahn echo measurements, we use an acousto-optical modulator.

The RF signals in this paper were generated by an Agilent E8257C source, whose output was gated using an RF switch (MiniCircuits ZASWA-2-50DR+). These signals were then combined, amplified to peak powers as high as 25 W (Amplifier Research 25S1G4A), and then sent to wiring in the cryostat. The RF and optical pulses were gated with pulse patterns generated by a digital delay generator (Stanford Research Systems DG645) and an arbitrary waveform generator (Tektronix AWG520). The phase of the Rohde & Schwartz signal was also controlled by the AWG520 through IQ modulation.

We used lock-in techniques to take all of the Hahn echo data in this paper. Specifically, we alternated the phase of the final $\pi/2$ microwave pulse of the Hahn echo sequence between $+\pi/2$ and $-\pi/2$. This alternation causes the spin coherence, at the end of the Hahn echo sequence, to be projected alternatively to opposite poles of the $m_s = +1$ / $m_s = +0$ Bloch sphere. Because the (*kk*)-divacancy's PL from the $m_s = +1$ pole of the Bloch sphere is stronger than that from the $m_s = +0$ pole, this alternation induces a change in PL ($\Delta$PL) between the two pulse sequences. Without spectrally filtering the PL, the ODMR contrast ($\Delta$PL / PL) is roughly 0.5%. When spectrally filtering the PL (which we did not do in this work), the ODMR contrast is 20% for the (*kk*)-divacancy. To transform the $\Delta$PL signals to a spin coherence measurement, we simply normalized the $\Delta$PL – $t_{\text{free}}$ traces, by dividing them by the maximum of the $\Delta$PL trace.

**Theoretical methods**

To calculate the Hahn-echo coherence of the (*kk*)-divacancy in 4*H*-SiC and the NV center in diamond, we considered a central spin model in which an electron spin with total spin 1 is coupled to an interacting nuclear spin bath through the secular electron-nuclear hyperfine interaction. Given the dilute nature of the nuclear spin density both in 4*H*-SiC (4.7% of $^{29}$Si and 1.1% of $^{13}$C) and diamond (1.1% of $^{13}$C), we only considered the direct dipole-dipole interaction for the nuclear-nuclear spin coupling. We calculated the full time-evolution of the combined qubit and nuclear bath system and computed the off-diagonal elements of the reduced qubit density matrix by tracing out the bath degrees of freedom at the end of the Hahn echo sequence [$\pi/2$ pulse − $t_{\text{free}}/2$ − $\pi$ pulse − $t_{\text{free}}/2$ − echo]. We considered randomly generated nuclear spin bath ensembles. A heterogeneous nuclear spin bath in 4*H*-SiC has around 1500 nuclear spins within 5 nm from the divacancy site, while the nuclear spin bath of diamond has around 1000 nuclear spins within 5 nm form the NV center. We used the cluster correlation expansion theory to systematically approximate the coherence function. Further details are found in Supplementary Notes 1-3.

**Code availability.** The codes that were used in this study are available upon request to the corresponding author.

**Data availability.** The data that support the findings of this study are available upon request to the corresponding author.

# References

1. Gruber, A., Drabenstedt, A., Tietz, C., Fleury, L., Wrachtrup, J. & Borczyskowski, von, C. scanning confocal optical microscopy and magnetic resonance on single defect centers. *Science* **276,** 2012–2014 (1997).
2. Jelezko, F., Gaebel, T., Popa, I., Domhan, M., Gruber, A. & Wrachtrup, J. Observation of Coherent Oscillation of a Single Nuclear Spin and Realization of a Two-Qubit Conditional Quantum Gate. *Phys. Rev. Lett.* **93,** 130501 (2004).
3. Zwanenburg, F. A., Dzurak, A. S., Morello, A., Simmons, M. Y., Hollenberg, L. C. L., Klimeck, G., Rogge, S., Coppersmith, S. N. & Eriksson, M. A. Silicon quantum electronics. *Rev. Mod. Phys.* **85,** 961–1019 (2013).
4. George, R. E., Edwards, J. P. & Ardavan, A. Coherent Spin Control by Electrical Manipulation of the Magnetic Anisotropy. *Phys. Rev. Lett.* **110,** 027601 (2013).
5. Xia, K., Kolesov, R., Wang, Y., Siyushev, P., Reuter, R., Kornher, T., Kukharchyk, N., Wieck, A. D., Villa, B., Yang, S. & Wrachtrup, J. O. All-Optical Preparation of Coherent Dark States of a Single Rare Earth Ion Spin in a Crystal. *Phys. Rev. Lett.* **115,** 093602 (2015).
6. DiVincenzo, D. P. The Physical Implementation of Quantum Computation. *Fortschritte der Physik* **48,** 771–783 (2000).
7. Schuster, D. I., Sears, A. P., Ginossar, E., DiCarlo, L., Frunzio, L., Morton, J. J. L., Wu, H., Briggs, G. A. D., Buckley, B. B., Awschalom, D. D. & Schoelkopf, R. J. High-Cooperativity Coupling of Electron-Spin Ensembles to Superconducting Cavities. *Phys. Rev. Lett.* **105,** 140501 (2010).
8. Hong, S., Grinolds, M. S., Pham, L. M., Le Sage, D., Luan, L., Walsworth, R. L. & Yacoby, A. Nanoscale magnetometry with NV centers in diamond. *MRS Bulletin* **38,** 155–161 (2013).
9. Rondin, L., Tetienne, J.-P., Hingant, T., Roch, J. F., Maletinsky, P. & Jacques, V. Magnetometry with nitrogen-vacancy defects in diamond. *Reports on Progress in Physics* **77,** 056503 (2014).
10. de Sousa, R. & Das Sarma, S. Theory of nuclear-induced spectral diffusion: Spin decoherence of phosphorus donors in Si and GaAs quantum dots. *Phys. Rev. B* **68,** 115322 (2003).
11. Mizuochi, N., Neumann, P., Rempp, F., Beck, J., Jacques, V., Siyushev, P., Nakamura, K., Twitchen, D. J., Watanabe, H., Yamasaki, S., Jelezko, F. & Wrachtrup, J. Coherence of single spins coupled to a nuclear spin bath of varying density. *Phys. Rev. B* **80,** 041201 (2009).
12. Balasubramanian, G., Neumann, P., Twitchen, D., Markham, M., Kolesov, R., Mizuochi, N., Isoya, J., Achard, J., Beck, J., Tissler, J., Jacques, V., Hemmer, P. R., Jelezko, F. & Wrachtrup, J. Ultralong spin coherence time in isotopically engineered diamond. *Nat. Mater.* **8,** 383–387 (2009).
13. Schweiger, A. & Jeschke, G. *Principles of Pulse Electron Paramagnetic Resonance*. (Oxford University Press, 2001).
14. Stanwix, P. L., Pham, L. M., Maze, J. R., Le Sage, D., Yeung, T. K., Cappellaro, P., Hemmer, P. R., Yacoby, A., Lukin, M. D. & Walsworth, R. L. Coherence of nitrogen-vacancy electronic spin ensembles in diamond. *Phys. Rev. B* **82,** 201201(R) (2010).
15. Tyryshkin, A. M., Morton, J. J. L., Benjamin, S. C., Ardavan, A., Briggs, G. A. D., Ager, J. W. & Lyon, S. A. Coherence of spin qubits in silicon. *J. of Phys.: Cond. Matt.* **18,** S783–S794 (2006).
16. George, R. E., Witzel, W., Riemann, H., Abrosimov, N. V., Nötzel, N., Thewalt, M. L. W. & Morton, J. J. L. Electron Spin Coherence and Electron Nuclear Double Resonance of Bi Donors in Natural Si. *Phys. Rev. Lett.* **105,** 067601 (2010).
17. Morley, G. W., Warner, M., Stoneham, A. M., Greenland, P. T., van Tol, J., Kay, C. W. M. & Aeppli, G. The initialization and manipulation of quantum information stored in silicon by bismuth dopants. *Nat. Mater.* **9,** 725–729 (2010).
18. Maze, J. R., Taylor, J. M. & Lukin, M. D. Electron spin decoherence of single nitrogen-vacancy defects in diamond. *Phys. Rev. B* **78,** 094303 (2008).
19. Zhao, N., Ho, S.-W. & Liu, R.-B. Decoherence and dynamical decoupling control of nitrogen vacancy center electron spins in nuclear spin baths. *Phys. Rev. B* **85, 115303** (2012).
20. Doherty, M. W., Dolde, F., Fedder, H., Jelezko, F., Wrachtrup, J., Manson, N. B. & Hollenberg, L. C. L. Theory of the ground-state spin of the $NV^-$ center in diamond. *Phys. Rev. B* **85,** 205203 (2012).
21. Childress, L., Dutt, M. V. G., Taylor, J. M., Zibrov, A. S., Jelezko, F., Wrachtrup, J., Hemmer, P. R. & Lukin, M. D. Coherent Dynamics of Coupled Electron and Nuclear Spin Qubits in Diamond. *Science* **314,** 281–285 (2006).
22. Hall, L. T., Cole, J. H. & Hollenberg, L. C. L. Analytic solutions to the central-spin problem for nitrogen-vacancy centers in diamond. *Phys. Rev. B* **90,** 075201 (2014).
23. Witzel, W. M., de Sousa, R. & Das Sarma, S. Quantum theory of spectral-diffusion-induced electron spin decoherence. *Phys. Rev. B* **72,** 161306(R) (2005).
24. Abe, E., Tyryshkin, A. M., Tojo, S., Morton, J. J. L., Witzel, W. M., Fujimoto, A., Ager, J. W., Haller, E. E., Isoya, J., Lyon, S. A., Thewalt, M. L. W. & Itoh, K. M. Electron spin coherence of phosphorus donors in silicon: Effect of environmental nuclei. *Phys. Rev. B* **82,** 121201 (2010).
25. Tyryshkin, A. M., Tojo, S., Morton, J. J. L., Riemann, H., Abrosimov, N. V., Becker, P., Pohl, H.-J., Schenkel, T., Thewalt, M. L. W., Itoh, K. M. & Lyon, S. A. Electron spin coherence exceeding seconds in high-purity silicon. *Nat. Mater.* **11,** 143–147 (2011).
26. Ryan, C. A., Hodges, J. S. & Cory, D. G. Robust Decoupling Techniques to Extend Quantum Coherence in Diamond. *Phys. Rev. Lett.* **105,** 200402 (2010).
27. De Lange, G., Wang, Z. H., Riste, D., Dobrovitski, V. V. & Hanson, R. Universal Dynamical Decoupling of a Single Solid-State Spin from a Spin Bath. *Science* **330,** 60–63 (2010).

28. Ma, W.-L., Wolfowicz, G., Zhao, N., Li, S.-S., Morton, J. J. L. & Liu, R.-B. Uncovering many-body correlations in nanoscale nuclear spin baths by central spin decoherence. *Nature Comm.* **5,** 4822 (2014).
29. Wolfowicz, G., Tyryshkin, A. M., George, R. E., Riemann, H., Abrosimov, N. V., Becker, P., Pohl, H.-J., Thewalt, M. L. W., Lyon, S. A. & Morton, J. J. L. Atomic clock transitions in silicon-based spin qubits. *Nat. Nanotech.* **8,** 561–564 (2013).
30. Mohammady, M. H., Morley, G. W., Nazir, A. & Monteiro, T. S. Analysis of quantum coherence in bismuth-doped silicon: A system of strongly coupled spin qubits. *Phys. Rev. B* **85,** 094404 (2012).
31. Balian, S. J., Wolfowicz, G., Morton, J. J. L. & Monteiro, T. S. Quantum-bath-driven decoherence of mixed spin systems. *Phys. Rev. B* **89,** 045403 (2014).
32. Christle, D. J., Falk, A. L., Andrich, P., Klimov, P. V., Hassan, J. U., Son, N. T., Janzen, E., Ohshima, T. & Awschalom, D. D. Isolated electron spins in silicon carbide with millisecond coherence times. *Nat. Mater.* **14,** 160–163 (2014).
33. Baranov, P. G., Ilin, I. V., Mokhov, E. N., Muzafarova, M. V., Orlinskii, S. B. & Schmidt, J. EPR Identification of the Triplet Ground State and Photoinduced Population Inversion for a Si–C Divacancy in Silicon Carbide. *JETP Letters* **82,** 441–443 (2005).
34. Son, N. T., Carlsson, P., Hassan, J. U., Janzén, E., Umeda, T., Isoya, J., Gali, A., Bockstedte, M., Morishita, N., Ohshima, T. & Itoh, H. Divacancy in 4H-SiC. *Phys. Rev. Lett.* **96,** 055501 (2006).
35. Gali, A. Time-dependent density functional study on the excitation spectrum of point defects in semiconductors. *phys. stat. sol. (b)* **248,** 1337–1346 (2011).
36. Koehl, W. F., Buckley, B. B., Heremans, F. J., Calusine, G. & Awschalom, D. D. Room temperature coherent control of defect spin qubits in silicon carbide. *Nature* **479,** 84–87 (2011).
37. Falk, A. L., Buckley, B. B., Calusine, G., Koehl, W. F., Dobrovitski, V. V., Politi, A., Zorman, C. A., Feng, P. X. L. & Awschalom, D. D. Polytype control of spin qubits in silicon carbide. *Nature Comm.* **4,** 1819 (2013).
38. Klimov, P. V., Falk, A. L., Buckley, B. B. & Awschalom, D. D. Electrically Driven Spin Resonance in Silicon Carbide Color Centers. *Phys. Rev. Lett.* **112,** 087601 (2014).
39. Falk, A. L., Klimov, P. V., Buckley, B. B., Ivády, V., Abrikosov, I. A., Calusine, G., Koehl, W. F., Gali, A. & Awschalom, D. D. Electrically and Mechanically Tunable Electron Spins in Silicon Carbide Color Centers. *Phys. Rev. Lett.* **112,** 187601 (2014).
40. Falk, A. L., Klimov, P. V., Ivády, V., Szász, K., Christle, D. J., Koehl, W. F., Gali, A. & Awschalom, D. D. Optical Polarization of Nuclear Spins in Silicon Carbide. *Phys. Rev. Lett.* **114,** 247603 (2015).
41. Ivády, V., Szász, K., Falk, A. L., Klimov, P. V., Christle, D. J., Janzen, E., Abrikosov, I. A., Awschalom, D. D. & Gali, A. Theoretical model of dynamic spin polarization of nuclei coupled to paramagnetic point defects in diamond and silicon carbide. *Phys. Rev. B* **92,** 115206 (2015).
42. Klimov, P. V., Falk, A. L., Christle, D. J., Dobrovitski, V. V. & Awschalom, D. D. Quantum entanglement at ambient conditions in a macroscopic solid-state spin ensemble. *Sci. Adv.* **1,** e1501015 (2015).
43. Yang, L.-P., Burk, C., Widmann, M., Lee, S.-Y., Wrachtrup, J. & Zhao, N. Electron spin decoherence in silicon carbide nuclear spin bath. *Phys. Rev. B* **90,** 241203 (2014).
44. Soltamov, V. A., Soltamova, A. A., Baranov, P. G. & Proskuryakov, I. I. Room Temperature Coherent Spin Alignment of Silicon Vacancies in 4H- and 6H-SiC. *Phys. Rev. Lett.* **108,** 226402 (2012).
45. Baranov, P. G., Bundakova, A. P., Soltamova, A. A., Orlinskii, S. B., Borovykh, I. V., Zondervan, R., Verberk, R. & Schmidt, J. Silicon vacancy in SiC as a promising quantum system for single-defect and single-photon spectroscopy. *Phys. Rev. B* **83,** 125203 (2011).
46. Riedel, D., Fuchs, F., Kraus, H., V ath, S., Sperlich, A., Dyakonov, V., Soltamova, A. A., Baranov, P. G., Ilyin, V. A. & Astakhov, G. V. Resonant Addressing and Manipulation of Silicon Vacancy

Qubits in Silicon Carbide. *Phys. Rev. Lett.* **109,** 226402 (2012).
47. Kraus, H., Soltamov, V. A., Riedel, D., Väth, S., Fuchs, F., Sperlich, A., Baranov, P. G., Dyakonov, V. & Astakhov, G. V. Room-temperature quantum microwave emitters based on spin defects in silicon carbide. *Nat. Phys.* **10,** 157–162 (2013).
48. Carter, S. G., Soykal, Ö. O., Dev, P., Economou, S. E. & Glaser, E. R. Spin coherence and echo modulation of the silicon vacancy in 4H-SiC at room temperature. *Phys. Rev. B* **92,** 161202 (2015).
49. Soykal, Ö. O., Dev, P. & Economou, S. E. Silicon vacancy center in 4H-SiC: Electronic structure and spin-photon interfaces. *Phys. Rev. B* **93,** 081207 (2016).
50. Kraus, H., Soltamov, V. A., Fuchs, F., Simin, D., Sperlich, A., Baranov, P. G., Astakhov, G. V. & Dyakonov, V. Magnetic field and temperature sensing with atomic-scale spin defects in silicon carbide. *Sci. Rep.* **4,** 5303 (2014).
51. Yang, W. & Liu, R.-B. Quantum many-body theory of qubit decoherence in a finite-size spin bath. *Phys. Rev. B* **78,** 085315 (2008).
52. Witzel, W. M. & Das Sarma, S. Quantum theory for electron spin decoherence induced by nuclear spin dynamics in semiconductor quantum computer architectures: Spectral diffusion of localized electron spins in the nuclear solid-state environment. *Phys. Rev. B* **74,** 035322 (2006).
53. Widmann, M., Lee, S.-Y., Rendler, T., Son, N. T., Fedder, H., Paik, S., Yang, L.-P., Zhao, N., Yang, S., Booker, I., Denisenko, A., Jamali, M., Momenzadeh, S. A., Gerhardt, I., Ohshima, T., Gali, A., Janzen, E. & Wrachtrup, J. Coherent control of single spins in silicon carbide at room temperature. *Nat. Mater.* **14,** 164–168 (2015).
54. Breuer, H. P. & Petruccione, F. *The Theory of Open Quantum Systems*. (OUP Oxford, 2007).
55. Van Oort, E. & Glasbeek, M. Optically detected low field electron spin echo envelope modulations of fluorescent N-V centers in diamond. *Chemical Physics* **143,** 131–140 (1990).
56. Ivanov, I. G., Yazdanfar, M., Lundqvist, B., Chen, J.-T., Ul-Hassan, J., Stenberg, P., Liljedahl, R., Son, N. T., Ager, J. W. I., Kordina, O. & Janzen, E. High-Resolution Raman and Luminescence Spectroscopy of Isotope-Pure $^{28}$Si$^{12}$C, Natural and $^{13}$C - Enriched 4H-SIC. *Materials Science Forum* **778-780,** 471–474 (2014).
57. Simin, D., Soltamov, V. A., Poshakinskiy, A. V., Anisimov, A. N., Babunts, R. A., Tolmachev, D. O., Mokhov, E. N., Trupke, M., Tarasenko, S. A., Sperlich, A., Baranov, P. G., Dyakonov, V. & Astakhov, G. V. All-optical dc nanotesla magnetometry using silicon vacancy fine structure in isotopically purified silicon carbide. *Phys. Rev. X* **6**, 031014 (2016).
58. Witzel, W. M., Carroll, M. S., Morello, A., Cywiński, Ł. & Das Sarma, S. Electron Spin Decoherence in Isotope-Enriched Silicon. *Phys. Rev. Lett.* **105,** 187602 (2010).
59. Weber, J. R., Koehl, W. F., Varley, J. B., Janotti, A., Buckley, B. B., Van de Walle, C. G. & Awschalom, D. D. Quantum computing with defects. *PNAS* **107,** 8513–8518 (2010).
60. Koehl, W. F., Seo, H., Galli, G. & Awschalom, D. D. Designing defect spins for wafer-scale quantum technologies. *MRS Bulletin* **40,** 1146–1153 (2015).
61. Szász, K., Ivády, V., Abrikosov, I. A., Janzén, E., Bockstedte, M. & Gali, A. Spin and photophysics of carbon-antisite vacancy defect in 4H silicon carbide: A potential quantum bit. *Phys. Rev. B* **91,** 121201 (2015).
62. Seo, H., Govoni, M. & Galli, G. Design of defect spins in piezoelectric aluminum nitride for solid-state hybrid quantum technologies. *Sci. Rep.* **6,** 20803 (2016).

# Acknowledgements

H.S. thank Nan Zhao and Setrak Balian for helpful discussions. H.S. is primarily supported by the National Science Foundation (NSF) through the University of Chicago MRSEC under award number DMR-1420709. GG is supported by DOE grant No. DE-FG02-06ER46262. D.D.A. was supported by the U.S. Department of Energy, Office of Science, Office of Basic Energy Sciences, Materials Sciences and Engineering Division. We acknowledge the University of Chicago Research Computing Center for support of this work. This work was supported by Air Force Office of Scientific Research (AFOSR), AFOSR-MURI, Army Research Office (ARO), NSF, and NSF-MRSEC.

## Author contributions

H.S. developed the numerical simulations and performed the theoretical calculations. A.L.F., P.V.K., and K.C.M. performed the optical experiments. D.D.A. and G.G. supervised the project. All authors contributed to the data analysis and production of the manuscript.

## Additional information

**Supplementary information** accompanies this paper.

**Competing financial interests:** The authors declare no competing financial interests.

**Correspondence and requests for materials** should be addressed to D.D.A. (email: awsch@uchicago.edu).

## Figure legends

**Figure 1. Defect spin qubits in nuclear spin baths. (a)** A depiction of the neutral ($kk$)-divacancy defect complex in 4$H$-SiC, in which a carbon vacancy ($V_C$, white sphere) at a quasi-cubic site ($k$) is paired with a silicon vacancy ($V_{Si}$, white sphere) formed at the nearest neighboring ($k$) site. **(b)** A depiction of the negatively charged NV center in diamond, which consists of a carbon vacancy ($V_C$, white sphere) paired with a substitutional nitrogen impurity (N, green sphere). Both defects have the same $C_{3v}$ symmetry (denoted by a grey pyramid) and spin-1 (black arrow) triplet ground state mainly derived from the surrounding carbon sp$^3$ dangling bonds. While the NV center spin is coupled to a homogeneous $^{13}$C nuclear spin bath (1.1%, $I_C$ = 1/2 represented with red arrows), the divacancy spin interacts with a heterogeneous nuclear spin bath of $^{13}$C and $^{29}$Si (4.7%, $I_{Si}$ = 1/2 represented with green arrows).

**Figure 2. Hahn-echo coherence of the divacancy ensemble in 4$H$-SiC. (a,b)** Experimental (a) and theoretical (b) Hahn-echo coherence of the $m_s$=+1 to $m_s$=0 ground-state spin transition of the divacancy ensemble with the $c$-axis-oriented magnetic field ($B$) at three different values. The experimental data was taken at $T$ = 20 K. **(c,d)** Experimental (c) and theoretical (d) Hahn-echo coherence of the spin transition

from (a) and (b), respectively as a continuous function of free evolution time ($t_{\text{free}}$) and $B$. The early loss of coherence near 47 mT in (c) corresponds to the spin triplet's ground state level anti-crossing (GSLAC).

**Figure 3. Analysis of the divacancy coherence.** (**a**) Experimental Hahn-echo coherence time ($T_2$) of the divacancy spin ensemble as a function of magnetic field ($B$) (filled circles) compared to theoretical $T_2$ of the divacancy (empty circles) and theoretical $T_2$ of the NV center in diamond (empty diamonds). The divacancy $T_2$ rises significantly, up to about 20 mT, and is then roughly constant, except for a dip at 47 mT, corresponding to the ground state level anti-crossing (GSLAC). (**b**) A direct comparison between the theoretical (red curve) and experimental (black curve) Hahn-echo coherence of the divacancy spin ensemble at two different magnetic fields of 17.5 mT (up) and 12.5 mT (down). (**c,d**) Experimental (c) and theoretical (d) Fast Fourier transform (FFT) power spectrum of the $m_s$=+1 to $m_s$=0 ground-state spin coherence data of the divacancy from Fig. 2 (c) and 2 (d), respectively. The frequency axis ($x$ axis) is normalized to $B$, so that the nuclear precession frequencies appear as vertical lines. Harmonics of these frequencies can also be seen both in theory and experiment. After 7 mT, the FFT intensities diminish as $B$ is increased. The hyperbolic features denoted by dotted arrows correspond to weak hyperfine interactions.

**Figure 4. Effective decoupling of the $^{13}$C and $^{29}$Si spin baths in 4$H$-SiC.** (**a**) The theoretical Hahn-echo coherence function of the divacancy ensemble at $B$ = 30 mT, calculated by only including the single- and heterogeneous pair-correlation contributions as defined in Eq. (3) and by varying the gyromagnetic ratio of $^{29}$Si ($\gamma_{\text{Si}}$) as a theoretical parameter while that of $^{13}$C ($\gamma_{\text{C}}$) is fixed at its experimental value. (**b**) The average number of homogeneous nuclear spin pairs whose lengths are less than 6 Å, as a function of distance from the divacancy qubit in 4$H$-SiC and from the NV center in diamond. The center-of-mass of a nuclear spin pair is used to measure the distance from the qubit. (**c**) The spatial distribution of homogeneous nuclear spin pairs in 4$H$-SiC and in diamond. The shortest homogeneous nuclear spin pair in diamond is 1.54 Å, corresponding to the C-C bond length, while that of the homogeneous nuclear spin pair in 4$H$-SiC is 3.07 Å, which is the second nearest neighboring Si-Si or C-C distances.

**Figure 5. Divacancy coherence time in isotopically purified 4$H$-SiC.** (**a,b,c,d,e,f**) Theoretical Hahn-echo coherence times ($T_2$) of the divacancy ensemble in 4$H$-SiC (a-e) and the NV center in diamond (f) as a function of $^{13}$C isotope concentration with a fixed $^{29}$Si concentration at 4.7 % (a), 3.0 % (b), 2.0% (c), 1.0 % (d), and 0.0 % (e) at five different magnetic fields. The black dashed line is the scaling law in Eq. (6) in the main text.

# Figures

**Figure 1.**

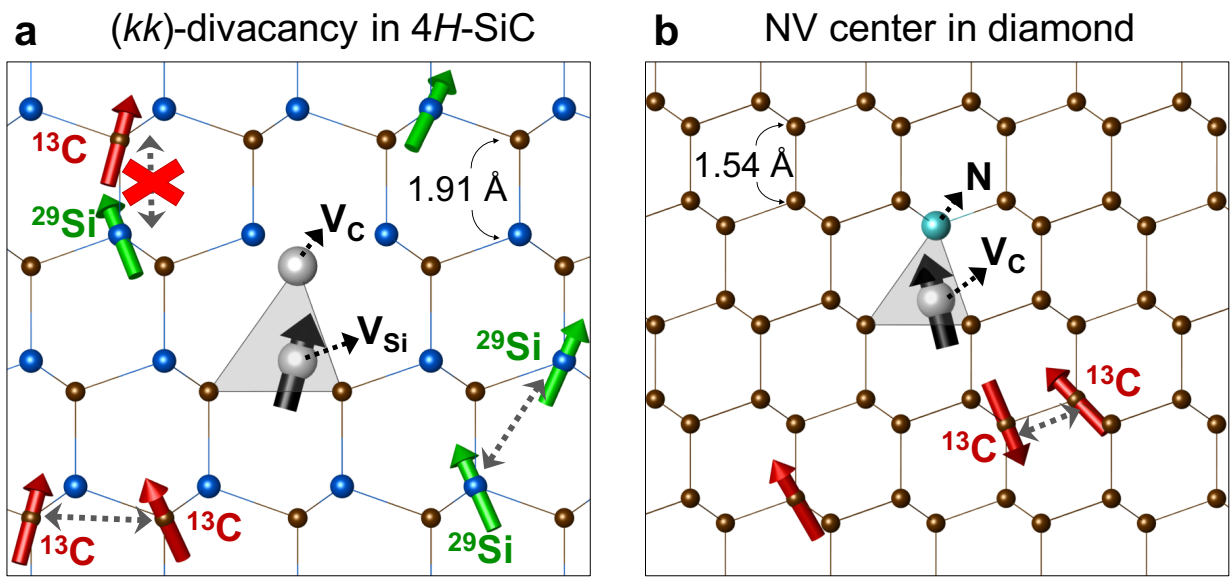

**Figure 2.**

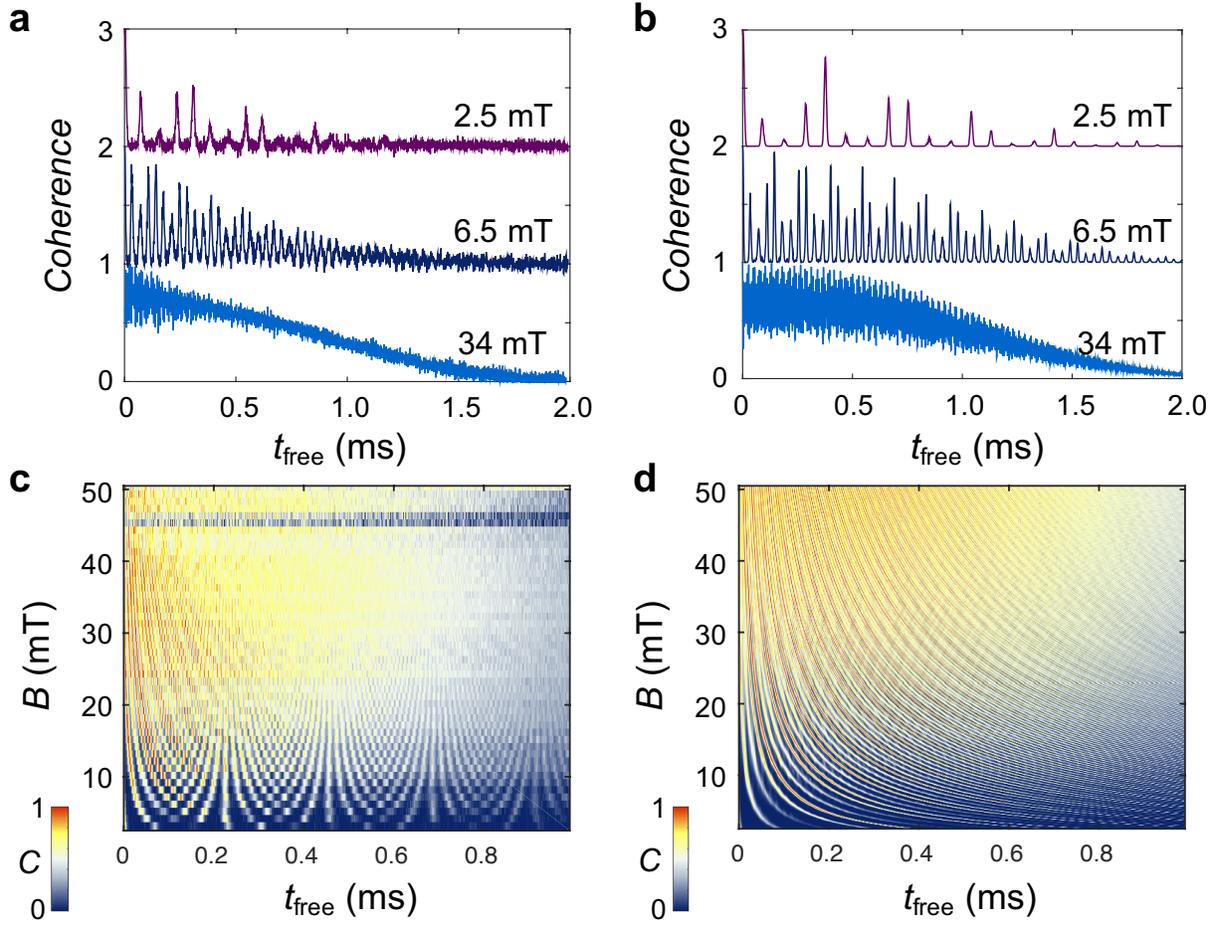

**Figure 3.**

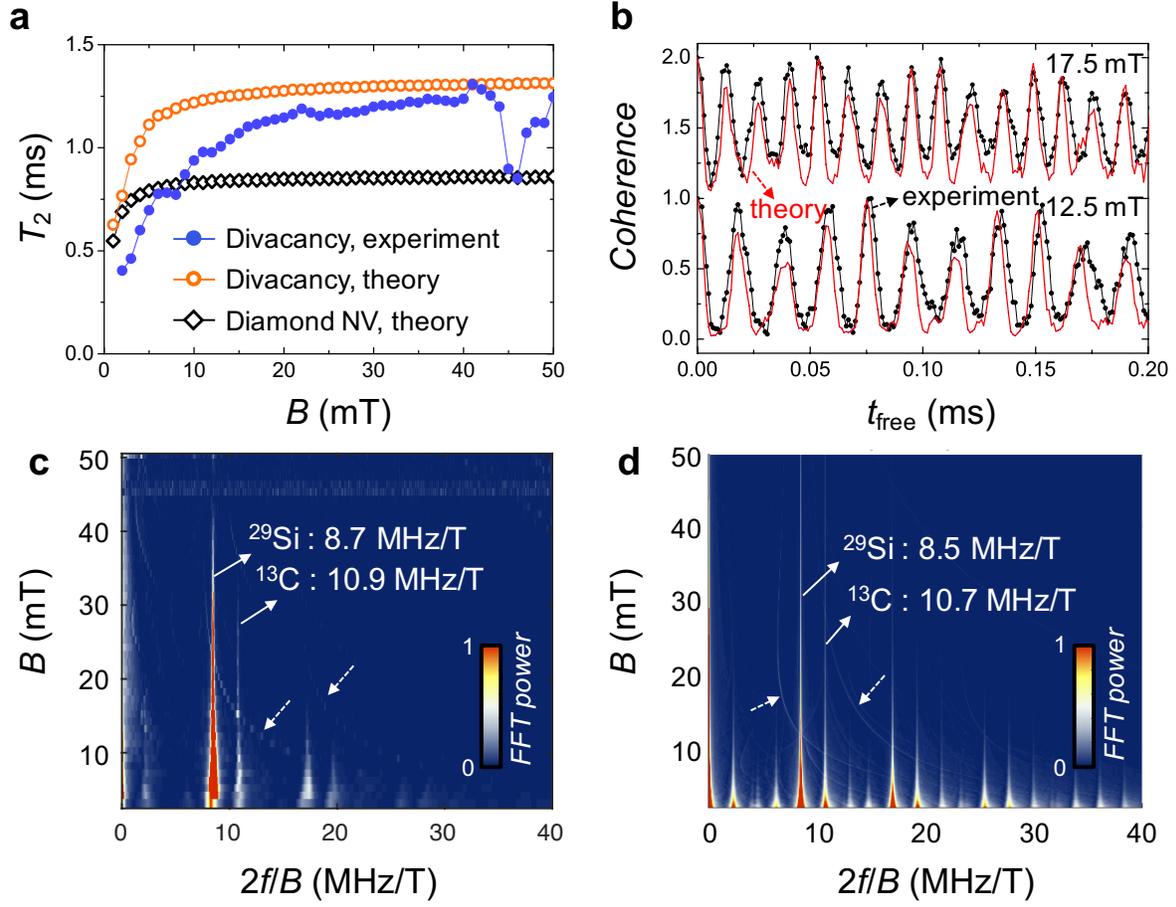

**Figure 4.**

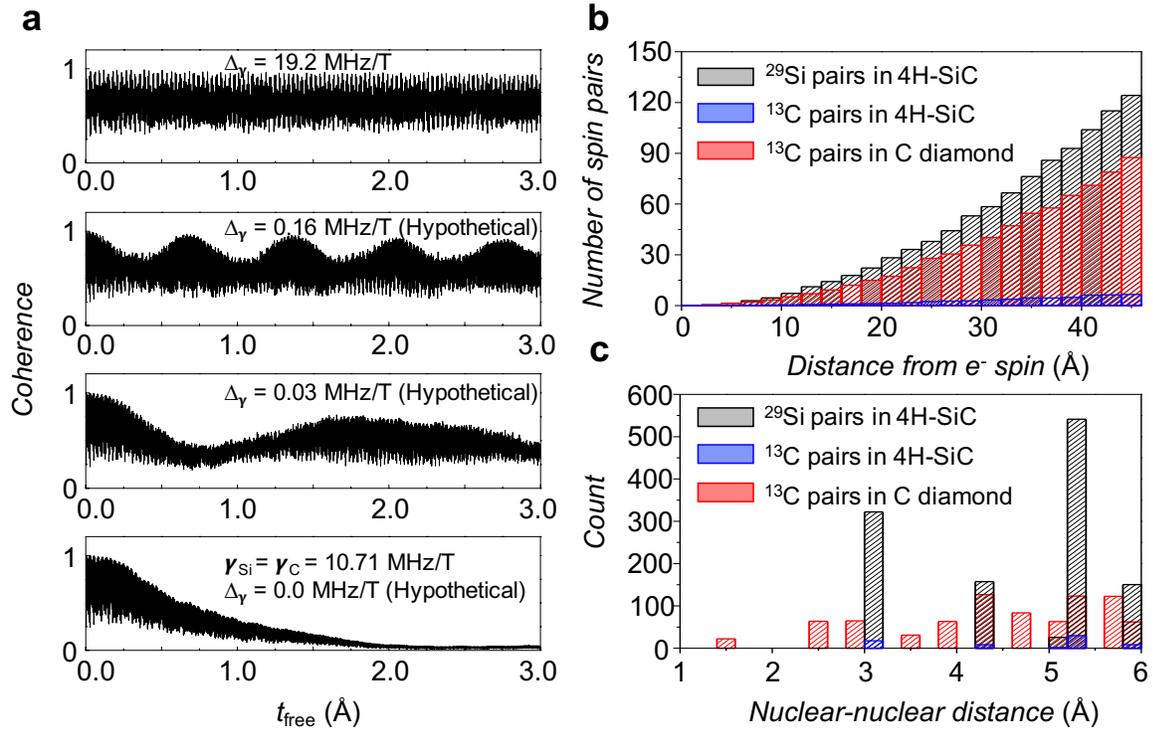

**Figure 5.**

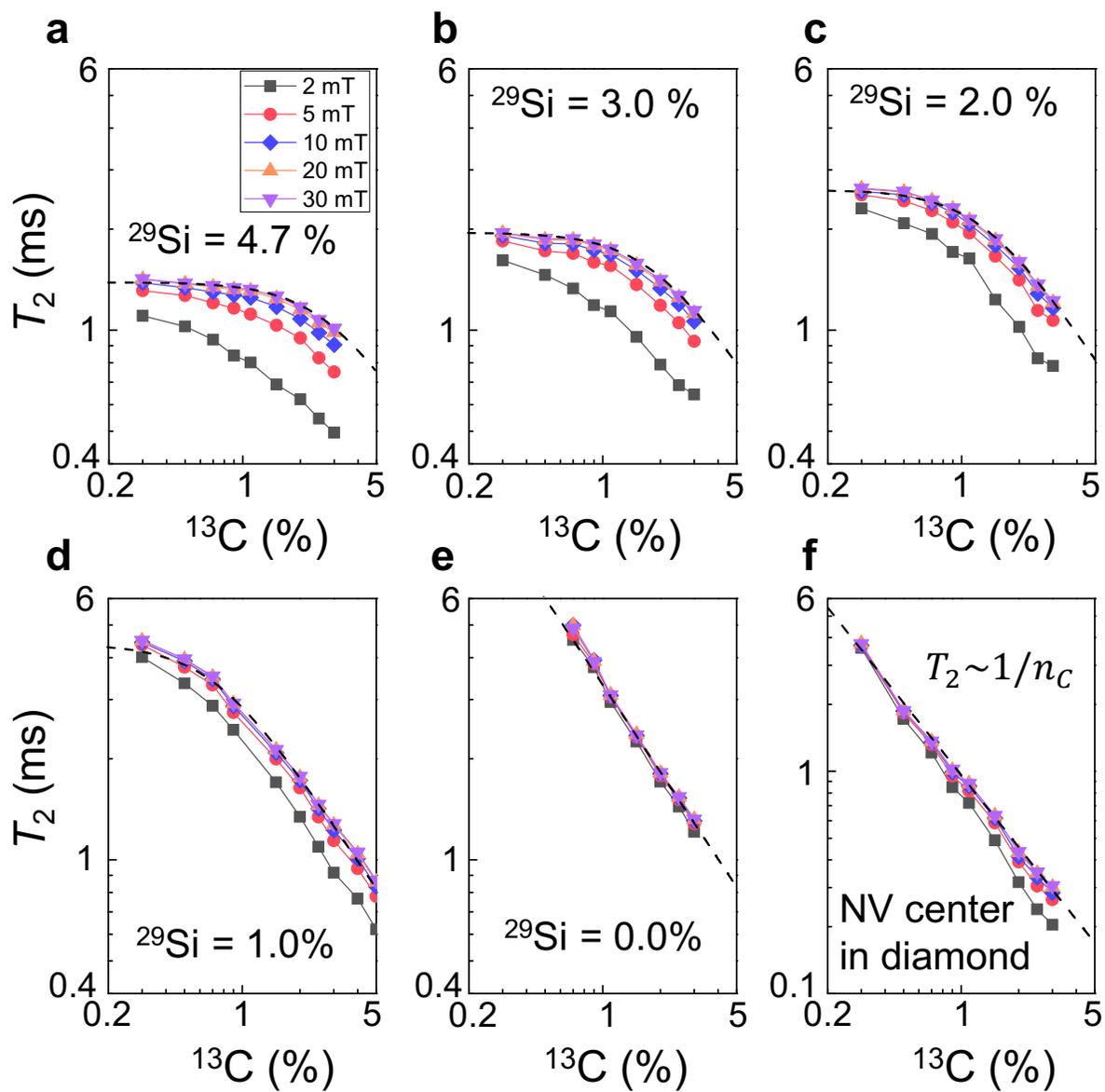

# Supplementary Information

## Supplementary Figures

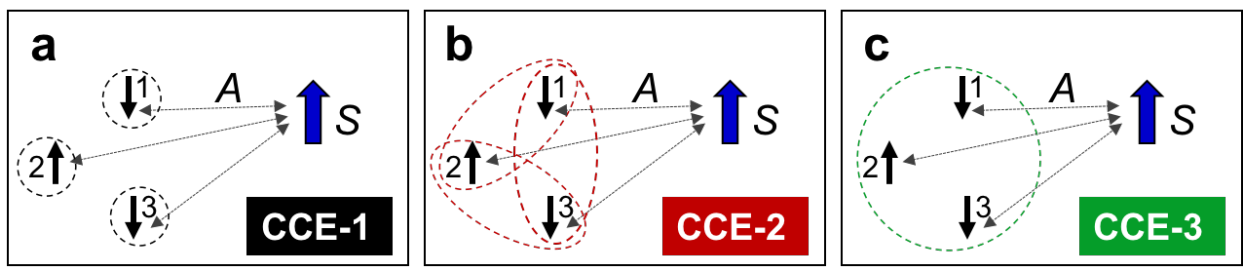

**Supplementary Figure 1. CCE method**. A system of electron spin ($S$) interacting with three nuclear spins (1,2,3) is considered. In CCE-1 (a), each nuclear spin is treated independently and it only interacts with the electron spin ($S$) through the hyperfine coupling ($A$). In CCE-2 (b) and CCE-3 (c), irreducible pair and triple correlations (see text) from possible nuclear spin pairs and triples, respectively, are recursively added to the single-correlation terms calculated in CCE-1. As there are only three nuclear spins in the bath, CCE-3 provides an exact solution of the electron spin coherence.

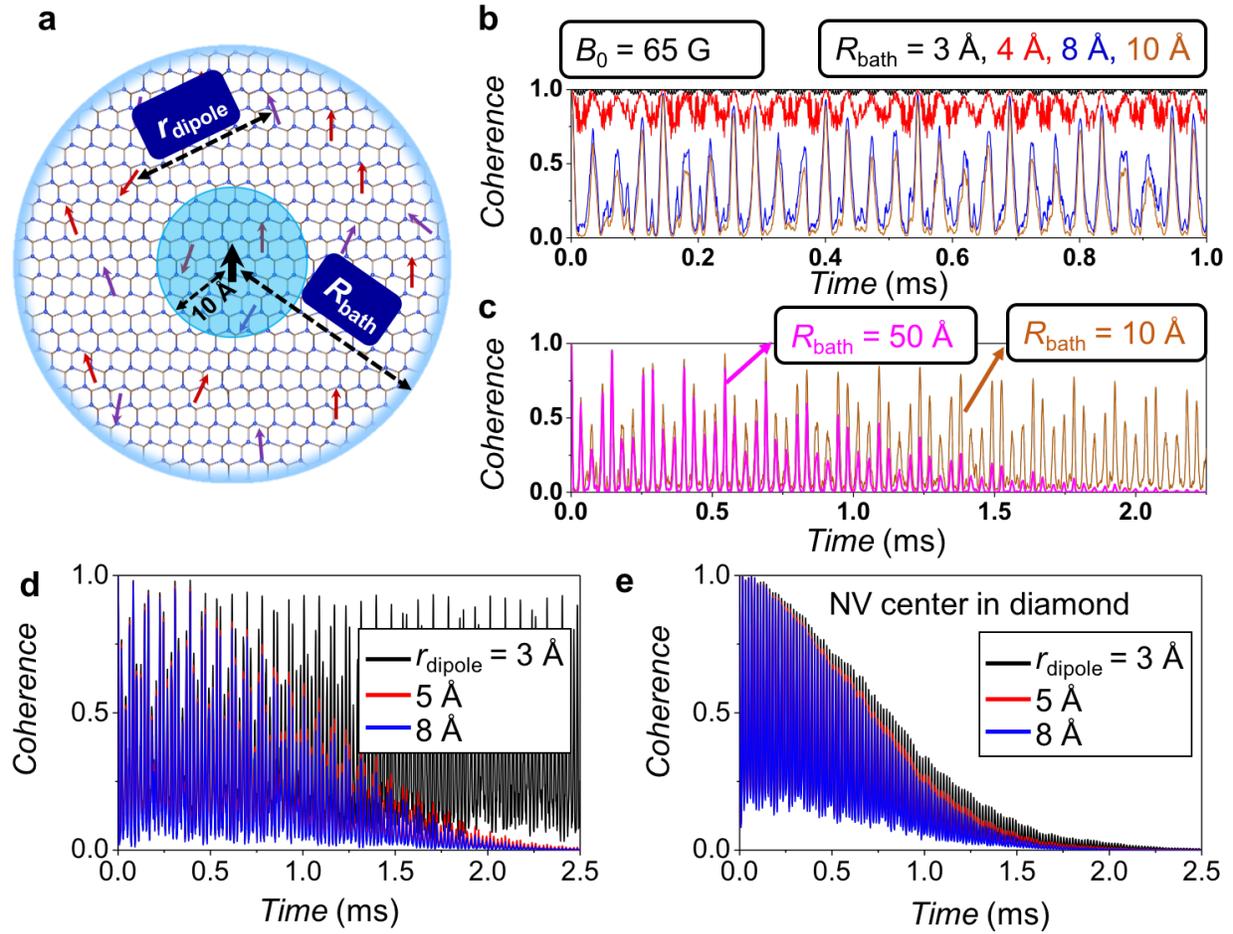

**Supplementary Figure 2. Numerical convergence tests of the Hahn-echo coherence.** (a) Schematic of a divacancy spin qubit (black arrow in the middle) coupled to a heterogeneous nuclear spin bath in 4*H*-SiC. Red arrows represent $^{29}$Si nuclear spins (4.7%, $I_{Si}$ = ½), while $^{13}$C nuclear spins (1.1%, $I_C$ = ½) are denoted by purple arrows. Two numerical parameters, $R_{bath}$ and $r_{dipole}$ are a cutoff radius for defining the bath size and a cutoff distance for the dipolar coupling between two nuclear spins, respectively. (b) The divacancy coherence at a magnetic field of 65 G at the CCE-2 level of theory calculated for four different bath sizes: black for $R_{bath}$ = 3 Å, red for 4 Å, blue for 8 Å, and brown for 10 Å. The coherence oscillation is mainly determined by nuclear spins within $R_{bath}$ = 10 Å, defining a strong coupling regime schematically shown as a blue area in (a). (c) The Hahn-echo coherence as in (b), but for $R_{bath}$ = 50 Å and 10 Å, showing that nuclear spins beyond $R_{bath}$ = 10 Å are mainly responsible for the coherence decay. The coherence function is numerically converged with $R_{bath}$ = 50 Å. (d), (e), The coherence of the divacancy (d) and the NV center (e) at a magnetic field of 115 G calculated for three different $r_{dipole}$.

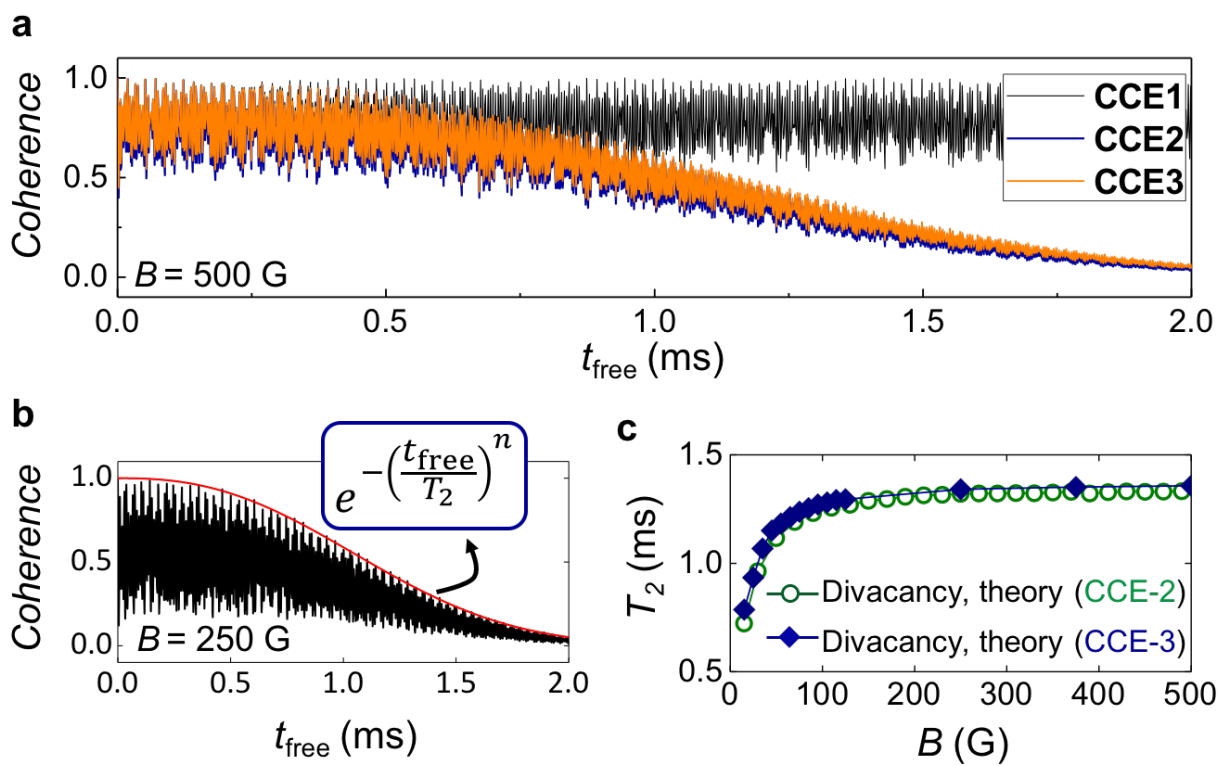

**Supplementary Figure 3. Numerical convergence of the coherence with respect to the CCE order.** (a) The coherence of the divancay at a magnetic field of 500 G calculated at the CCE-1 (black), CCE-2 (blue), and CCE-3 (orange) levels of theory. CCE-3 does not give significant correction to the CCE-2 results, implying that the CCE-2 approximation provides numerically converged results. (b) Fitting of the divacancy coherence at $B = 250$ G with a stretched exponential function having two parameters: the Hahn-echo coherence time $T_2$ and an stretching exponent $n$. (c) $T_2$ of the divacancy as a function of static magnetic field at the CCE-2 and CCE-3 levels of theory. The CCE-2 and CCE-3 results are in excellent agreement with each other, further providing the numerical validity of CCE-2.

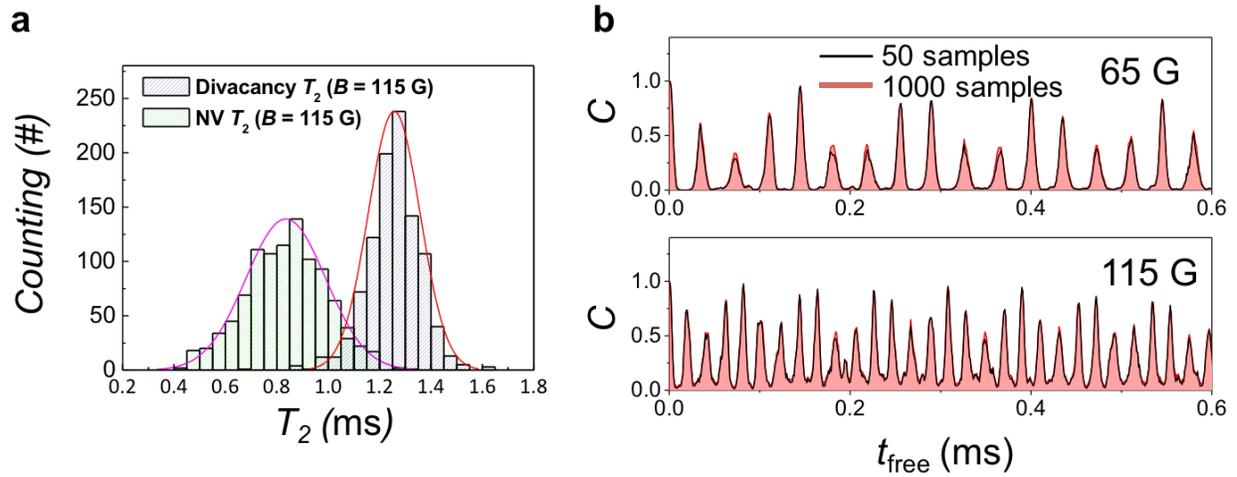

**Supplementary Figure 4. Ensemble statistics**. (a) Distribution of $T_2$ of the divacancy ensemble in 4*H*-SiC and the NV ensemble in diamond at a static magnetic field of 115 G. Red curves are normal distribution fit of the histograms. (b) Direct comparison of the coherence of the divacancy in 4*H*-SiC averaged over 50 different random nuclear spin baths (Black curve) to the coherence averaged over 1000 nuclear spin baths (Filled red curve) at $B = 65$ G (up) and $B = 115$ G (down).

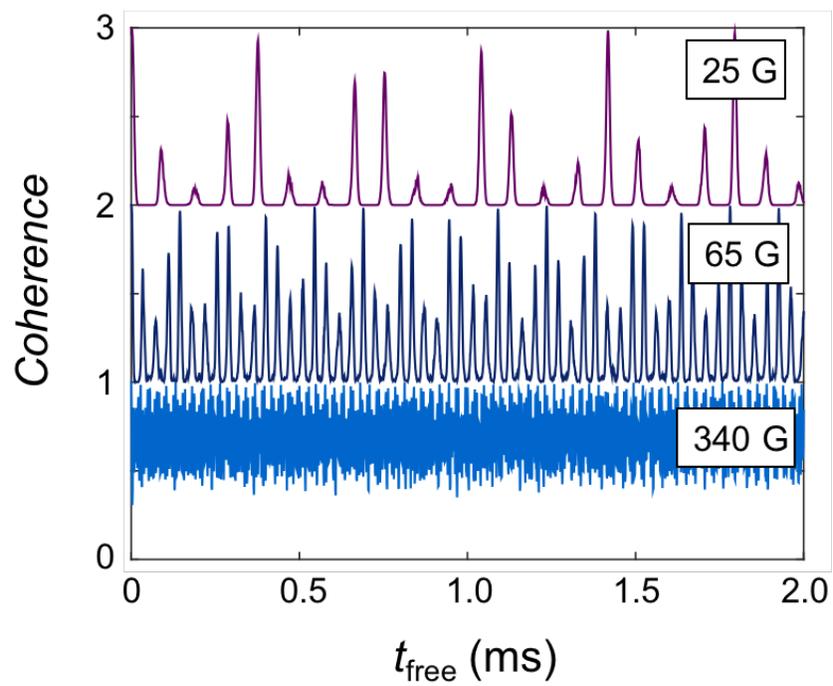

**Supplementary Figure 5. ESEEM spectra calculated within CCE-1**. Analytical expression in Supplementary Equation 16 is used along with the same numerical strategy used for the results in Figure 2 (b) in the main article. The CCE-1 calculations reproduce all the features in the coherence functions in Figure 2 (b) of the main article except for the overall envelop decay.

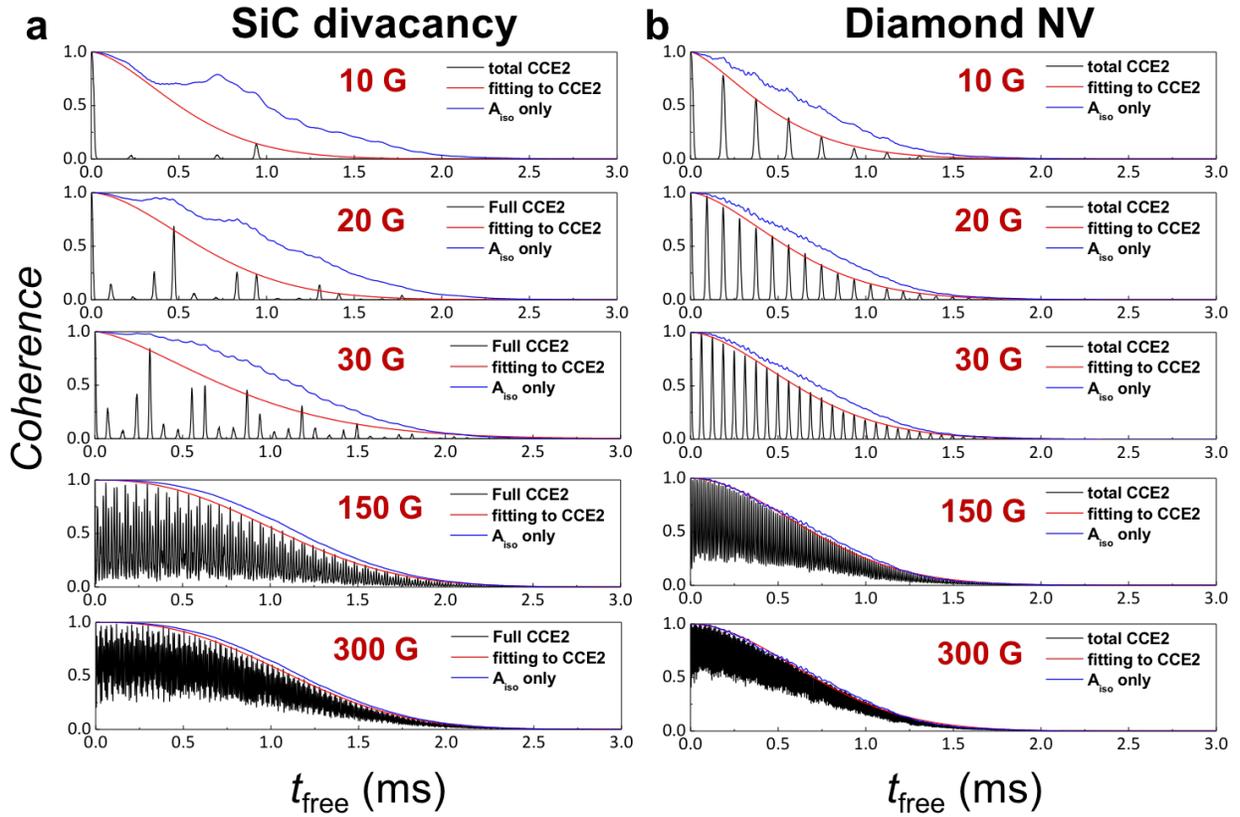

**Supplementary Figure 6. Pseudo-secular hyperfine field induced coherence decay.** The Hahn-echo coherence of the divacancy in 4$H$-SiC (a) and the NV center in diamond (b) at several magnetic fields calculated with the full hyperfine coupling ($A_i$, $B_{ix}$, and $B_{iy}$ in Supplementary Equation 7) (black curve) and without anisotropic hyperfine coupling (blue curve, $A_i$ only). The red curve is a fit to the full CCE-2 curve and the difference between the red curve and the $A_i$-only blue curve is the contribution from the pseudo-secular hyperfine interactions to the coherence decay, which becomes negligible at a large magnetic field beyond 100 G for both divacancy and NV.

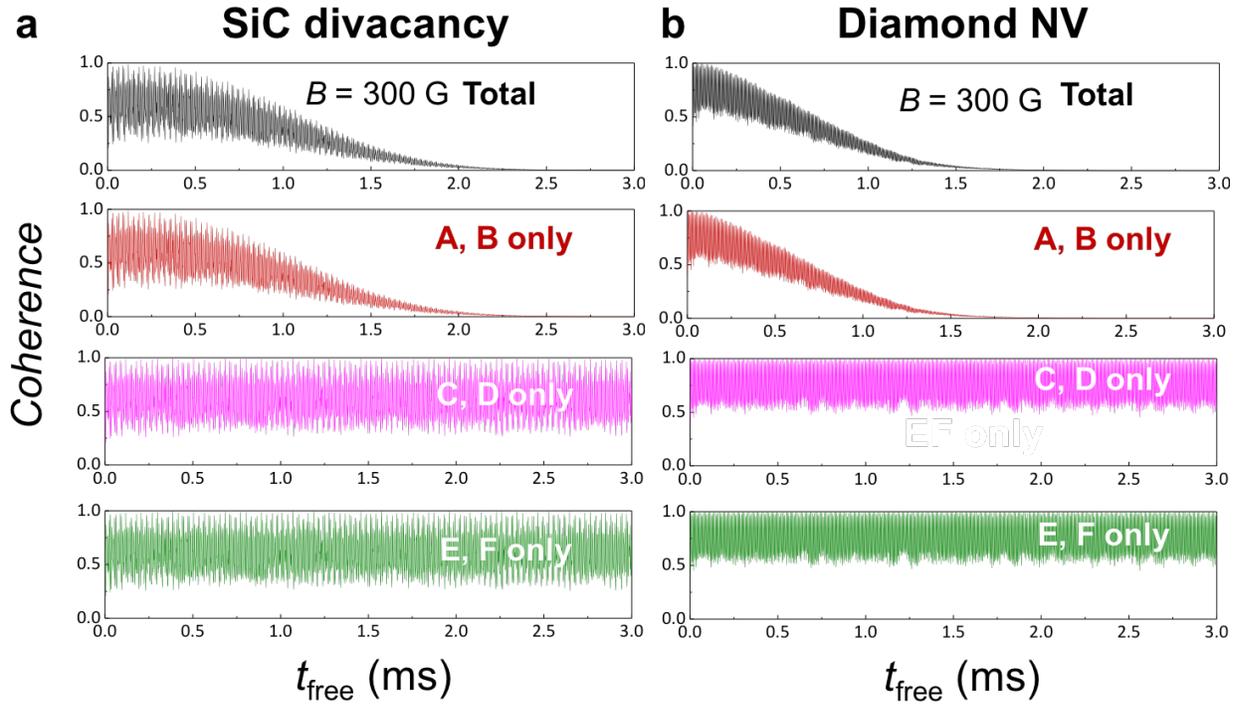

**Supplementary Figure 7. Central spin decoherence induced by nuclear spin flip-flop transitions.** The full Hahn-echo coherence of the divacancy in 4*H*-SiC (a) and the NV center in diamond (b) at $B = 300$ G at the top (black curve) compared to those calculated only with the *A* and *B* term (red curve), the *C* and *D* terms (cyan curve) and the *E* and *F* terms (green curve) of the nuclear dipole-dipole coupling Hamiltonian shown in Supplementary Equation 18. At a large magnetic field above $B = 100$ G, the nuclear flip-flop transitions induced by the *A* and *B* terms are the main cause of the coherence decay for both divacancy and NV qubits, while the transitions induced by the other terms are fully suppressed.

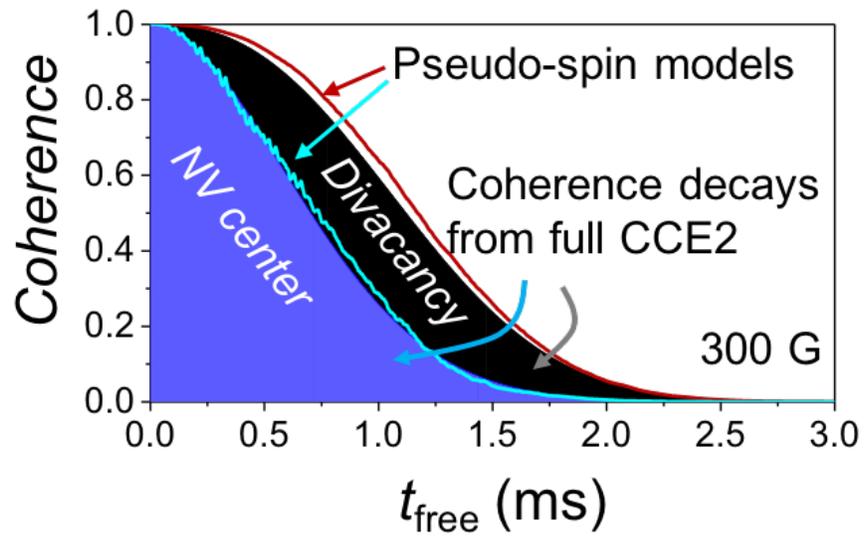

**Supplementary Figure 8. Pseudo-spin model of decoherence.** The coherence decay of the divacancy in 4*H*-SiC (filled black curve) and the NV center in diamond (filled blue curve) from the full CCE calculations, for which only the envelop decay is shown for clarity. The red and the cyan curves are the coherence decay curves of the divacancy and the NV center, respectively, calculated by using the pseudo-spin model shown in Supplementary Equation 23.

# Supplementary Tables

| Defect spin qubit | $B_0$ (G) | Number of samples | $T_2$ average (ms) | $T_2$ STDEV ($N$-1) (ms) | $n$ average (ms) | $n$ STDEV ($N$-1) (ms) |
|---|---|---|---|---|---|---|
| ($kk$)-divacancy in 4$H$-SiC | 65 | 50 | 1.17 | 0.18 | 2.26 | 0.37 |
| | | 100 | 1.19 | 0.15 | 2.30 | 0.32 |
| | | 1000 | 1.18 | 0.14 | 2.27 | 0.32 |
| | 115 | 50 | 1.26 | 0.11 | 2.42 | 0.33 |
| | | 100 | 1.27 | 0.10 | 2.47 | 0.26 |
| | | 1000 | 1.26 | 0.10 | 2.45 | 0.25 |
| NV center in diamond | 65 | 50 | 0.816 | 0.182 | 2.178 | 0.492 |
| | | 100 | 0.807 | 0.168 | 2.222 | 0.487 |
| | | 1000 | 0.796 | 0.166 | 2.220 | 0.464 |

**Supplementary Table 1.** Computed ensemble-averaged $T_2$ and $n$ of the divacancy qubit in 4$H$-SiC and the NV center in diamond along with their standard deviation (STDEV) computed with the ($N$-1) method, where $N$ is the number of samples in ensemble.

## Supplementary Note 1. Theoretical calculations of decoherence dynamics

**Quantum bath approach to qubit decoherence.**

To calculate the decoherence dynamics of divacancy spin qubits in 4*H*-SiC, we use a microscopic quantum bath approach, in which a combined qubit and bath system is considered as a closed quantum system[1]. The phase information of a qubit at an arbitrary time *t* is encoded in the off-diagonal element of the reduced density matrix, for which the bath degrees of freedom are traced out. Suppose that a combined qubit and bath system is initialized at *t*=0 as a product state as follows:

$$|\Psi(0)\rangle = \frac{1}{\sqrt{2}}(|1\rangle + |0\rangle) \otimes |\mathcal{B}(0)\rangle, \qquad (7)$$

where $|1\rangle$ and $|0\rangle$ are up and down states of the qubit, respectively, and $|\mathcal{B}(0)\rangle$ is an initial state of the bath. In the course of time evolution, the bath state may be entangled with the qubit state:

$$|\Psi(\tau)\rangle = \frac{1}{\sqrt{2}}\left(|0\rangle \otimes |\mathcal{B}^{(0)}(\tau)\rangle + |1\rangle \otimes |\mathcal{B}^{(1)}(\tau)\rangle\right). \qquad (8)$$

The off-diagonal element of the reduced density matrix is then given as an overlap between the two bath states ($|\mathcal{B}^{(0)}(\tau)\rangle$ and $|\mathcal{B}^{(1)}(\tau)\rangle$). Therefore, in order to use the quantum-bath method, we need to identify the dominant bath degrees of freedom of a given system and calculate the bath evolution conditioned on qubit states.

It has been established for the nitrogen-vacancy (NV) center in diamond that the main source of the spin decoherence is its coupling to the $^{13}$C nuclear spin bath (1.1% abundance, $I_C = 1/2$) and other paramagnetic defect centers such as N impurities (P1 centers) in the lattice[2]. The later can be controlled by a chemical purification process and the longest Hahn-echo ensemble coherence time ($T_2$) of the NV centers in high-purity diamond has been measured to be 0.63 ms[3]. A similar argument can be applied to the divacancy qubits in 4*H*-SiC except that the nuclear spin bath of 4*H*-SiC is a heterogeneous one having both naturally occurring $^{29}$Si isotopes (4.7%, $I_{Si}=1/2$) and $^{13}$C isotopes. Other paramagnetic defects might be generated during sample preparation. We note, however, that a defect density in our samples is very low as described in the main text. The divacancy density is approximately $10^{12}$ cm$^{-3}$ [4] and an unintentional dopant density is also very low ($5 \times 10^{13}$ cm$^{-3}$)[5]. Considering a paramagnetic defect density of $10^{13} \sim 10^{14}$ cm$^{-3}$, there may be one or two paramagnetic impurities within 1000 ~ 3000 Å from a divacancy qubit in 4*H*-SiC with dipolar coupling strengths ranging from 50 Hz to 2 Hz, while there are already ~ 10 nuclear spins even within 10 Å with electron-nuclear dipolar coupling strengths ranging from 0.1 MHz to 0.01 MHz. Thus, contribution from paramagnetic defect centers to the divacancy decoherence may be negligible in our SiC samples and we only focus on the effect of the fluctuating nuclear spin bath due to the nuclear-nuclear dipolar interactions.

It is worth discussing about possible temperature effects on nuclear and electron spins in SiC and diamond. In principles, electron and nuclear spins can be randomly flipped at a finite temperature, inducing qubit decohernece[6]. Temperature-induced flipping of nuclear and electron spins can be characterized by nuclear and electronic spin-lattice relaxation times, $T_{1,n}$ and $T_{1,e}$, respectively. It has been found that the $T_{1,n}$ times in SiC and diamond are extremely long exceeding several hours owing to the lack of efficient nuclear spin-lattice coupling mechanism[7,8]. The time scale of the NV center coherence and that of the divacancy qubits has been measured to be ~ millisecond, meaning the $T_{1,n}$-induced nuclear spin flipping to be negligible in this time scale. $T_{1,e}$-induced relaxation of a central electron qubit maybe another issue in SiC and diamond at a finite temperature as the qubit's $T_2$ time is ultimately limited by $2T_{1,e}$[2,9]. Temperature-dependent $T_{1,e}$ times of the NV center and the divacancy have been measured to be ranging from $6 \times 10^{-3}$ s (at T = 300 K) to $2 \times 10^2$ s (at T = 10 K)[10] and from $6.2 \times 10^{-4}$ s (T = 250 K) to $2.0 \times 10^{-2}$ s (T = 20 K)[4], respectively. Therefore, we also ignore the $T_{1,e}$-induced relaxation effect on the central qubit decoherence in SiC and diamond at T = 20 K.

**Spin Hamiltonian and Hahn-echo coherence function.**

Considering the fluctuating nuclear spin bath as a main source of the divacancy decoherence, we can write down the spin Hamiltonian as $\mathcal{H}_{total} = \mathcal{H}_S + \mathcal{H}_B + \mathcal{H}_{S-B}$, where $\mathcal{H}_S$ and $\mathcal{H}_B$ are terms for the qubit and the bath under a static magnetic field ($\vec{B} = B_0 \hat{z}$), respectively, while $\mathcal{H}_{S-B}$ accounts for the hyperfine coupling between the qubit and the bath[9]. Each term can be written as follows:

$$\mathcal{H}_S = -\gamma_e \hbar \vec{B} \cdot \vec{S} + \Delta S_z^2, \tag{9}$$

$$\mathcal{H}_B = -\vec{B} \cdot \sum_i \gamma_i \hbar \vec{I}_i + \mathcal{H}_{n-n}, \tag{10}$$

$$\mathcal{H}_{int} = \vec{S} \cdot \sum_i \vec{A}_i \cdot \vec{I}_i, \tag{11}$$

where $\gamma_e$ and $\gamma_i$ ($i$ = C or Si) are the gyromagnetic ratios of electron and nuclear spins of $^{29}$Si and $^{13}$C isotopes, respectively, and they are given as $\gamma_e$= -1.761 × 10$^{11}$ rad s$^{-1}$ T$^{-1}$, $\gamma_{Si}$= -5.319 × 10$^7$ rad s$^{-1}$ T$^{-1}$ and $\gamma_C$= 6.728 × 10$^7$ rad s$^{-1}$ T$^{-1}$. The second term in $\mathcal{H}_S$ is the zero-field splitting tensor splitting the $m_s$=0 and $m_s$=±1 sublevels of the electron spin ($S$ = 1) and it has been measured to be 1.305 GHz for the *(kk)*-divacancy spin in 4*H*-SiC[11]. $\mathcal{H}_{n-n}$ is the magnetic dipole-dipole coupling between two nuclear spins and it is given by:

$$\mathcal{H}_{\text{n-n}} = \frac{\mu_0}{4\pi} \sum_{i<j} \gamma_i \gamma_j \hbar^2 \left( \frac{\vec{I}_i \cdot \vec{I}_j}{r_{ij}^3} - \frac{3(\vec{I}_i \cdot \vec{r}_{ij})(\vec{I}_j \cdot \vec{r}_{ij})}{r_{ij}^5} \right), \qquad (12)$$

where $r_{ij}$ is the distance between the nuclear spin $I_i$ and $I_j$. The hyperfine tensor ($\overleftrightarrow{A}_i$) that couples the electron spin to the $i$-th nuclear spin in the bath may have two parts: the isotropic Fermi contact interaction and the anisotropic dipole-dipole interaction[9]. The Fermi contact term is mainly derived from the overlap between the defect's electron spin density and the nuclear spin under consideration. We note, however, that the defect spin density is highly localized in space owing to the localized nature of the carbon $sp^3$ dangling bonds[11]. Thus, the Fermi contact term may become negligible compared to other energy scales in the Hamiltonian beyond three to four nearest neighboring sites. In this study, we ignore the Fermi contact term. In addition, we ignore the off-diagonal non-secular $S_x$ and $S_y$ terms in the anisotropic dipolar coupling because the zero-field splitting of GHz order of magnitude and the large difference between the electron and nuclear gyromagnetic ratios would make the hyperfine-induced flipping of the electron spin almost impossible in the time-scale that we are interested in. This 'secular approximation' is also a valid approximation if the spin-lattice relaxation time $T_{1,e}$ is much larger than the pure-dephasing time $T_2$[9], which is our case[4,10]. The final form of the hyperfine interaction is written as follows:

$$\mathcal{H}_{\text{int}} = S_z \sum_i \vec{A}_i \cdot \vec{I}_i = \sum_i (B_{ix} I_{ix} S_z + B_{iy} I_{iy} S_z + A_i I_{iz} S_z), \qquad (13)$$

where $\vec{A}_i$ is the hyperfine field for the $i^{\text{th}}$ nuclear spin $I_i$, consisting of secular $A_i$ hyperfine coupling and pseudo-secular $B_{ix}$ and $B_{iy}$ hyperfine couplings. We note that the hyperfine field is only active when the electron spin is not in the $m_s = 0$ state. We also observe that the secular coupling term gives rise to the Zeeman frequency shift for a nuclear spin while the pseudo-secular coupling terms can flip the nuclear spin, thus creating a fluctuation in the nuclear spin bath at low magnetic fields. Within the secular approximation, the total Hamiltonian commutes with the $S_z$ operator and the electron spin is preserved, allowing us to project the total Hamiltonian on the electron spin basis. As a result, we obtain the following pure-dephasing Hamiltonian[1]:

$$\mathcal{H}_{\text{total}} = \sum_{m_s=-1}^{+1} |m_s\rangle \langle m_s| \otimes \mathcal{H}_{m_s}, \qquad (14)$$

where $\mathcal{H}_{m_s}$ is the bath Hamiltonian conditioned on the electron spin sub-level $m_s$.

$$\mathcal{H}_{m_s} = \omega_{m_s} + \mathcal{H}_B + m_s \sum_i \vec{A}_i \cdot \vec{I}_i, \qquad (15)$$

where $\omega_{m_s}$ is the energy spectrum of the electron spin. We note that the same Hamiltonian and the same approximation are applied to the NV center in diamond except that the C lattice only has $^{13}$C nuclear spins and there is $^{14}$N-derived nuclear spin ($I_N$=1) associated with the NV center.

The coherence function, the off-diagonal element of the reduced density matrix, can be formally written as:

$$\mathcal{L}(t) \equiv \frac{tr[\rho_{tot}(t)S_+]}{tr[\rho_{tot}(0)S_+]}, \qquad (10)$$

where $S_+ = S_x + iS_y$ is the electron spin raising operator and $\rho_{tot}$ is the density of operator of the combined qubit ($\rho_S$) and bath ($\rho_B$) system. At $t = 0$, we assume that the system is initialized as the product state as $\rho_{tot}(0) = \rho_S(0) \otimes \rho_B(0)$ and it evolves in time as $\rho_{tot}(t) = \mathcal{U}(t)\rho_{tot}(0)\mathcal{U}^\dagger(t)$, where $\mathcal{U}(t)$ is the Hahn-echo propagator[9]. We employ the assumption of piecewise constant Hamiltonian, in which the Hahn-echo propagator in the rotating frame breaks into a $\pi/2$-pulse bringing the initial down-state ($m_s = 0$) into a superposition of the up ($m_s = +1$) and down states, followed by a free-evolution under a given static magnetic field for $t_{free}/2$, an ideal $\pi$-pulse ($P_\pi = -i\sigma_x$), and another $t_{free}/2$ free-evolution under static $B$-field, subsequently. Noting that the free evolution operator is block-diagonal as the pure-dephasing Hamiltonian in Supplementary Equation 8 does not mix the up and down states of the electron spin, one can finally write down the Hahn-echo coherence as:

$$\mathcal{L}(t_{free}) = tr[\rho_{tot}(t_{free})S_+] = tr_B[\mathcal{U}_-^\dagger \mathcal{U}_+^\dagger \mathcal{U}_- \mathcal{U}_+ \rho_{Bath}(0)] = \sum_\mathcal{J} \mathcal{P}_\mathcal{J} \langle \mathcal{J}|\mathcal{U}_-^\dagger \mathcal{U}_+^\dagger \mathcal{U}_- \mathcal{U}_+|\mathcal{J}\rangle, \qquad (11)$$

where $\rho_{Bath}(0) = \sum \mathcal{P}_\mathcal{J}|\mathcal{J}\rangle\langle\mathcal{J}|$. $\mathcal{U}_+ = e^{-(i/\hbar)(\mathcal{H}_B + \sum_i \vec{A}_i \cdot \vec{I}_i)t_{free}/2}$ and $\mathcal{U}_- = e^{-(i/\hbar)\mathcal{H}_B t_{free}/2}$ are free bath propagators conditioned on the up and down states of the electron spin, respectively. We note that at T = 20 K, the nuclear spin bath is almost completely thermalized, making the initial nuclear spin bath density matrix to be the identity.

## Supplementary Note 2. Cluster correlation expansion

**Concepts and numerical implementation.**

Supplementary Equation 11 formally allows for calculating the coherence of the divacancy and the NV qubits. However, the direct matrix calculations are still an unfeasible task as a large number of nuclear spins are involved. For instance, there are around 1500 nuclear spins in 4*H*-SiC and 1000 nuclear spins in diamond within 5 nm from a divacancy qubit and a NV center, respectively, leading to a matrix dimension of $2^{1000}$ to $2^{1500}$ to be solved. Recently developed cluster correlation expansion (CCE) technique[12,13] enables a systematic approximation to the coherence function. The basic concept of CCE is schematically shown in Supplementary Figure 1. Suppose a spin qubit is coupled to a bath of three nuclear spins. The simplest approximation is to ignore all the interactions between the nuclear spins and treat them independently, yielding a CCE-1 coherence function that is a product of all the 'single-correlation' terms as schematically shown in Supplementary Figure 1 (a).

$$\mathcal{L}_1(t_{\text{free}}) = \prod_i \tilde{\mathcal{L}}_i(t_{\text{free}}) = \prod_i \mathcal{L}_i(\tau)/\tilde{\mathcal{L}}_0, \qquad (12)$$

where $i$ is an index for nuclear spins ($i$=1,2,3) and $\tilde{\mathcal{L}}_0$ is a normalization constant or 'empty-correlation' term. Apparently, the independent nuclear spin model cannot capture dipole-dipole induced bath fluctuations[14]. The next-order approximation would be to include two-body or pair-correlation effects (see Supplementary Figure 1 (b)):

$$\mathcal{L}_2(t_{\text{free}}) = \prod_i \tilde{\mathcal{L}}_i(t_{\text{free}}) \prod_{\{i,j\}} \tilde{\mathcal{L}}_{i,j}, \qquad (13)$$

where $\tilde{\mathcal{L}}_{i,j} = \mathcal{L}_{i,j}(t_{\text{free}})/(\tilde{\mathcal{L}}_i \tilde{\mathcal{L}}_j)$. Note that if two nuclear spin pairs share one nuclear spin in common (see Supplementary Figure 1), the dipole-dipole induced transitions of the two pairs may be correlated to each other. This three-body correlation can be captured at the next CCE-3 level of theory:

$$\mathcal{L}_3(t_{\text{free}}) = \prod_i \tilde{\mathcal{L}}_i(t_{\text{free}}) \prod_{\{i,j\}} \tilde{\mathcal{L}}_{i,j} \prod_{\{i,j,k\}} \tilde{\mathcal{L}}_{i,j,k} \qquad (14)$$

where $\tilde{\mathcal{L}}_{i,j,k} = \mathcal{L}_{i,j,k}(t_{free})/(\tilde{\mathcal{L}}_i \tilde{\mathcal{L}}_j \tilde{\mathcal{L}}_k)/(\tilde{\mathcal{L}}_{i,j}\tilde{\mathcal{L}}_{j,k}\tilde{\mathcal{L}}_{i,k})$. In this simple example of the 3-nuclear-spin model, we remark that the CCE-3 coherence function in Supplementary Equation 14 is the same as the exact coherence function, i.e. $\mathcal{L}_3(t_{\text{free}}) = \mathcal{L}_{1,2,3}(t_{\text{free}})$. This means that for any possible nuclear spin baths, CCE expansion provides the exact solution when the expansion includes the largest possible nuclear spin clusters (i.e. the entire nuclear spin bath). For practical calculations, the expansion would stop at a certain cluster size *N*, and the CCE-*N* expansion is given by:

$$\mathcal{L}_N(\tau) = \prod_{C \subseteq \{1,2,3,\ldots,N\}} \tilde{\mathcal{L}}_C(\tau), \tag{15}$$

where all the irreducible cluster correlations up to clusters with $N$ nuclear spins being included. $N$ for a specific system can be determined by calculating the numerical convergence with respect to $N$, which will be further discussed later in this article.

We used C/C++ and the Eigen3 library[15] to implement the CCE method. We created orthorhombic supercells of 4$H$-SiC and C diamond and placed a ($kk$)-divacancy defect and a NV center in the middle of the SiC and C supercells, respectively. We used experimentally determined lattice structures of 4$H$-SiC and diamond and the $c$-direction of the supercells are aligned with the $C_{3v}$-axis of the defects: (0001) for the ($kk$)-divacancy and (111) for NV, along which static magnetic field is applied. The presence of nuclear spins in the lattices naturally occurring from $^{13}$C and $^{29}$Si isotopes were simulated by randomly placing $^{13}$C and $^{29}$Si nuclear spins at 1.1% and 4.7% concentrations in the supercells. The same strategy was used to generate multiple supercells for creating an ensemble of random heterogeneous nuclear spin baths of $^{29}$Si and $^{13}$C in 4$H$-SiC and an ensemble of homogeneous nuclear spin baths with $^{13}$C for C diamond. The size of the supercell and the number of supercells in an ensemble have been systematically determined by checking the numerical convergence with respect to the bath size and the ensemble average, which will be described in the next section.

**Numerical convergence.**

There are a number of numerical parameters that need to converge in our CCE calculations: (1) size of the nuclear spin bath ($R_{\text{bath}}$), (2) the largest dipole-dipole interaction distance between two nuclear spins ($r_{\text{dipole}}$), and (3) the CCE expansion order. In this section, we discuss each of them and their physical implications. All calculations done in this section are ensemble-averaged over 50 nuclear spin bath samples. Convergence of the ensemble average will be discussed in the next section. In addition, we only discuss results for the divacancy in 4$H$-SiC for simplicity. The convergence test results for NV in diamond will be discussed briefly at the end of this section.

In the supercell geometry discussed in 2-a, a central $S = 1$ spin qubit (either divacancy or NV) is coupled to a random nuclear spin bath mainly through the electron-nuclear dipolar coupling, which decays as $1/R^3$ where $R$ is the distance between the electron spin and a nuclear spin under consideration. Thus, beyond a certain cutoff radius defined as $R_{\text{bath}}$ the $e$-$n$ coupling may become negligible, defining the bath size as shown in Supplementary Figure 2 (a). In Supplementary Figure 2 (b), we calculate the divacancy coherence function for different bath sizes under a static magnetic field of 65 G at the CCE-2 level of theory. As noted in the main text the divacancy coherence function comprises of the electron spin echo envelop modulation (ESEEM) and the overall decay. In Supplementary Figure 2 (b), the ESSEM pattern rapidly emerges as the bath size increases from 3 Å (only including the nearest neighboring sites) to 8 Å. A further increase of the bath size to 10 Å does not significantly change the oscillation pattern, indicating that the origin of the ESSEM spectrum is the strong hyperfine coupling with ~ 10 nuclear spins within 10 Å. In addition, we find that nuclear spins beyond the strong coupling regime (See Supplementary Figure 2 (a)) is mainly responsible for the coherence decay as shown in Supplementary Figure 2 (c), which compares the divacancy coherence function calculated with the small bath of $R_{bath}$ = 10 Å to that of a larger nuclear spin bath of $R_{bath}$ = 50 Å. In addition, we note that the coherence function does not change as we vary the bath size from 40 Å to 60 Å. Therefore, we set $R_{bath}$ = 50 Å to be our cutoff radius for the nuclear spin bath that the central ($kk$)-divacancy is coupled with. This observation also lays down a solid ground for our quantum-bath approach to decoherence, which assumes that the combined qubit and bath system form a closed quantum system. Our numerical convergence tests show that this assumption is self-consistently valid for the ($kk$)-divacancy coupled with nuclear spins within $R_{bath}$ = 50 Å.

In principles, CCE calculations at a given expansion order, e.g. CCE-2, should involve all possible pairs of nuclear spins. However, some remote nuclear spins would not interact strong enough to contribute to the coherence decay because the nuclear dipole-dipole coupling scales as $1/r^3$, where $r$ is the distance between two nuclear spins. Thus, we introduced a cutoff distance, $r_{dipole}$ and we treat two nuclear spins as independent spins if they are separated by more than $r_{dipole}$. We perform CCE-2 calculations for various $r_{dipole}$ values and we found that the numerical convergence is achieved for $r_{dipole}$ = 6 Å and we used $r_{dipole}$ = 8 Å for all calculations for this work.

Practical CCE calculations are terminated at a certain CCE order known as the CCE-$N$ approximation, where $N$ indicates the number of nuclear spins in the largest cluster considered. The order of CCE calculations should depend on the problem under investigation and should be determined by checking the numerical convergence with respect to the CCE order. In Supplementary Figure 3 (a), we show representative coherence functions of the divacancy qubit calculated at difference CCE orders. We found that the CCE-2 and CCE-3 coherence functions show negligible differences, indicating that CCE-2 calculations provide full numerical convergence. We further verify the numerical convergence by comparing $T_2$ as a function of static magnetic field in Supplementary Figure 3 (c). We note that the CCE-2 and CCE-3 results of $T_2$ show negligible difference across a wide range of magnetic field.

The validity of the CCE-2 approximation on our problem could be understood by considering that our nuclear spin concentration in the lattice is very low and the nuclear dipole-dipole interaction decays fast as $1/r^3$. Given our $r_{dipole}$ of around 6 to 8 Å, it is hard to form a significant number of strongly coupled nuclear spin triples, but most of the nuclear spins would form either isolated spins or spin pairs whose pair-wise spin transitions are unlikely correlated[13,16].

We found the same numerical convergence behavior for the NV center in diamond. Therefore, we apply $R_{bath}$ of 50 Å and $r_{dipole}$ of 8 Å to all the divacancy and NV calculations.

**Statistics for ensemble averages.**

The quantum bath model described above suggests that the decoherence dynamics of a spin qubit coupled to a nuclear spin bath may significantly depend on the specific nuclear spin arrangement in a given bath, thus giving rise to variations in $T_2$ in an ensemble of random nuclear spin baths. Supplementary Figure 4 shows the histograms of $T_2$ (see Supplementary Figure 3 (b) for definition) of an (*kk*)-divacancy ensemble with 1000 different random nuclear spin baths at a static magnetic field of 115 G. $T_2$ shows significant variation across the nuclear samples, but eventually follows a normal distribution consistent with the central limit theorem. At a magnetic field of 115 G, the divacancy $T_2$ is centered around 1.3 ms, while some nuclear spin configurations give rise to 0.9 ms to 1.7 ms single spin coherence time.

To compare with experiments, we perform ensemble averages of the coherence functions and the $T_2$ times and we find that ensemble averages over 50 samples are good enough to produce numerically converged results. Supplementary Figure 4 (b) shows a direct comparison of the coherence function of the (*kk*)-divacancy spin ensemble averaged over 1000 samples to that averaged over 50 samples. We note that while the average over 1000 samples smooths out some minor noisy features on the coherence function, the overall shape is already well-converge with the average over 50 samples. Supplementary Table 1 summarizes ensemble $T_2$ and $n$ of the (*kk*)-divacancy in 4*H*-SiC and the NV center in diamond at two magnetic fields of 65 G and 115 G, showing that the average over 50 samples provides converged $T_2$ and $n$ for both systems.

# Supplementary Note 3. Analytic equations of Hahn-echo coherence

**Electron Spin Echo Envelop Modulation.**

One of the main features in the coherence described in the main text is a rapid collapse and revival as a function of free evolution time $t_{\text{free}}$, which is known as electron spin echo envelop modulation (ESEEM) in the literature[14,17]. As hinted by the FFT power spectrum analysis shown in Figure 3 in the main article, ESEEM is driven by single nuclear spin precessions, hence the main ESEEM feature can be captured at the CCE-1 level of theory, i.e. independent nuclear spin approximation. As no nuclear-nuclear interactions are present in CCE-1, one can analytically solve the coherence equation by using, e.g. the product operator formalism[9] and the solution is given as:

$$\mathcal{L}_{\text{CCE1}}(t_{\text{free}}) = \prod_i \left(1 - 2k_i \sin^2\left(\sqrt{(\omega_i + A_i)^2 + B_i^2} \frac{t_{\text{free}}}{4}\right) \sin^2\left(\omega_i \frac{t_{\text{free}}}{4}\right)\right), \quad (16)$$

where $i$ runs over all single nuclear spins in the bath, $\omega_i$ is the nuclear Larmor frequency, and $A_i$ and $B_i$ (= $(B_{ix}^2+B_{iy}^2)^{1/2}$) are secular and pseudo-secular hyperfine interactions, respectively. $k_i$ is called the modulation depth parameter[9], which is given as:

$$k_i = \frac{B_i^2}{(\omega_i + A_i)^2 + B_i^2}. \quad (17)$$

In Supplementary Figure 5, we calculate the coherence function at the three different magnetic fields from Figure 2 (b) in the main article by using the Supplementary Equation 16 and we note that the ESEEM oscillation is perfectly reproduced. When the applied static magnetic field increases the Larmor frequency increases, making the modulation depth parameter to go to zero. Therefore, the coherence oscillation amplitude is suppressed as observed in Supplementary Figure 5 and Figure 2 in the main article.

**Pseudo-spin models of spin qubit decoherence.**

To understand the decoherence dynamics of the (*kk*)-divacancy spin compared to the NV decoherence, we employ a pseudo-spin model, which has been applied to the NV center in the literature[18,19]. To check the applicability of the pseudo-spin model, we determine the most important Hamiltonian terms for the coherence decay. In Supplementary Figure 6, we calculate the coherence function of the divacancy in 4*H*-SiC and the NV center in diamond only with secular hyperfine interactions ($A_i$ in Supplementary Equation 7) and compare it to the full CCE-2 calculation results. We note that the $A_i$-only calculations lacks the ESEEM feature (see Supplementary Equation 16 and 17 for the reason), while it captures the decay behavior especially for a magnetic field larger than 100 G for both NV and divacancy. For small magnetic fields under 100 G, there is significant contribution from the pseudo-secular hyperfine interactions as they can effectively flip the nuclear spins owing to the small Zeeman splitting, inducing significant spin fluctuation in the bath. However, as the magnetic field increases more than $B = 100$ G, the Zeeman splitting increases and the pseudo-secular hyperfine induced nuclear spin flipping is suppressed, making the secular approximation for the hyperfine coupling good enough to describe the coherence decay.

Nuclear spins in diamond and 4*H*-SiC interact with each other by the nuclear dipole-dipole interaction (Supplementary Equation 6), inducing pairwise nuclear spin transitions. This can be easily seen by rewriting the dipolar Hamiltonian between nuclear spin n1 and n2 in Supplementary Equation 6 as follows[20]:

$$\mathcal{H}_{n1-n2} = \frac{\mu_0}{4\pi} \frac{\gamma_{n1}\gamma_{n2}\hbar^2}{r_{12}^3} (A + B + C + D + E + F), \tag{18}$$

where

$$
\begin{aligned}
A &= I_{1z}I_{2z}(1 - 3\cos^2\theta), \\
B &= -\frac{1}{4}(I_{1+}I_{2-} + I_{1-}I_{2+})(1 - 3\cos^2\theta), \\
C &= -\frac{3}{2}(I_{1+}I_{2z} + I_{1z}I_{2+})\sin\theta\cos\theta\, e^{-i\phi}, \\
D &= -\frac{3}{2}(I_{1-}I_{2z} + I_{1z}I_{2-})\sin\theta\cos\theta\, e^{+i\phi}, \\
E &= -\frac{3}{4}I_{1+}I_{2+}\sin^2\theta\, e^{-2i\phi}, \\
F &= -\frac{3}{4}I_{1-}I_{2-}\sin^2\theta\, e^{+2i\phi}.
\end{aligned}
\tag{19}
$$

In Supplementary Figure 7, we calculate the Hahn-echo coherence of both divacancy and NV only with the *AB*, *CD*, or *EF* terms and compare them to the full Hahn-echo coherence function in order to identify the most important pairwise nuclear spin transitions. We find that at a magnetic field larger than 100 G, e.g. $B = 300$ G, CCE-2 calculations only keeping the nuclear spin flip-flop *AB* terms reproduce the full CCE-2 result, while similar calculations only using the *CD* or *EF* terms do not induce any coherence decay. For the NV center in diamond, the ↑↑ and ↓↓ configurations are well separated in energy from each other and from the ↑↓ and ↓↑ states due to the large Zeeman splitting, thus only the *AB* flip-flop transitions become the main pairwise transitions[18,19]. For the flip-flop transition, the ↑↓ and ↓↑ states are separated in energy by the difference in the hyperfine fields imposed by the electron spin shown in Supplementary Equation 7. For 4*H*-SiC, as explained in the main article, all possible pair-wise transitions for heterogeneous nuclear spin pairs are fully suppressed, thus only pairwise transitions in homogeneous spin pairs, e.g. either $^{13}$C – $^{13}$C or $^{29}$Si – $^{29}$Si, remain active for the coherence decay. Therefore, the same flip-flop *AB* terms in the dipole-dipole coupling becomes the most important interaction channels even for the heterogeneous nuclear spin bath in 4*H*-SiC.

With the observations made so far (Supplementary Figure 6 and Supplementary Figure 7), we can construct a pseudo-spin model for a homogeneous nuclear spin pair interacting with a spin qubit by keeping only the secular hyperfine term ($A_i$) and the flip-flop term from the dipole-dipole interaction. The Hilbert space for the pseudo-spin model only contains the two ↑↓ and ↑↓ nuclear spin states and the pseudo-spin Hamiltonian (for nuclear spin 1 and 2) can be written as:

$$\mathcal{H}_{12}^{m_s} = D_{12}^{m_s} J_x + \Omega_{12}^{m_s} J_z, \qquad (20)$$

where $m_s$ is the electron spin sub-level (either 0 or 1 for NV and divacancy), and $J_z$ and $J_x$ are fictious spin-1/2 operators. $\Omega_{12}^{m_s}$ is a pseudo-spin frequency depending on the electron spin sublevel and in our case, it is given as:

$$\Omega_{12}^{m_s=+1} = \Delta A_{12} = A_1 - A_2,$$
$$\Omega_{12}^{m_s=0} = 0. \qquad (21)$$

$D_{12}^{m_s}$ is a pseudo-spin transition rate conditioned on the electron spin state, derived from the secular nuclear dipole-dipole interaction:

$$D_{12}^{m_s=+1} = D_{12}^{m_s=0} \equiv D_{12} = \frac{1}{2}\left(\frac{\mu_0}{4\pi}\frac{\gamma_1 \gamma_2 \hbar^2}{r_{12}^3}\right)(3\cos^2\theta_{12} - 1). \qquad (22)$$

Then, the Hahn-echo coherence function of the divacancy spin (or the NV center) coupled to this homogeneous nuclear spin pair is given as:

$$\mathcal{L}_{\text{pair}}(t_{\text{free}}) = 1 - K_{12}\sin^2\left(\sqrt{(\Delta A_{12})^2 + D_{12}^2}\,\frac{t_{\text{free}}}{4}\right)\sin^2\left(D_{12}\frac{t_{\text{free}}}{4}\right), \qquad (23)$$

where

$$K_{12} = \frac{(\Delta A_{12})^2}{(\Delta A_{12})^2 + D_{12}^2}. \qquad (24)$$

Supplementary Figure 8 compares the coherence function calculated using the pseudo-spin model to the coherence decay from the full CCE calculation for both the NV and divacancy defects. We observe that the pseudo-spin model reproduces the overall coherence decay well for both NV and divacancy.

## Supplementary References

1. Breuer, H. P. & Petruccione, F. *The Theory of Open Quantum Systems*. (OUP Oxford, 2007).
2. Balasubramanian, G., Neumann, P., Twitchen, D., Markham, M., Kolesov, R., Mizuochi, N., Isoya, J., Achard, J., Beck, J., Tissler, J., Jacques, V., Hemmer, P. R., Jelezko, F. & Wrachtrup, J. Ultralong spin coherence time in isotopically engineered diamond. *Nat. Mater.* **8,** 383–387 (2009).
3. Stanwix, P. L., Pham, L. M., Maze, J. R., Le Sage, D., Yeung, T. K., Cappellaro, P., Hemmer, P. R., Yacoby, A., Lukin, M. D. & Walsworth, R. L. Coherence of nitrogen-vacancy electronic spin ensembles in diamond. *Phys. Rev. B* **82,** 201201(R) (2010).
4. Falk, A. L., Buckley, B. B., Calusine, G., Koehl, W. F., Dobrovitski, V. V., Politi, A., Zorman, C. A., Feng, P. X. L. & Awschalom, D. D. Polytype control of spin qubits in silicon carbide. *Nature Comm.* **4,** 1819 (2013).
5. Christle, D. J., Falk, A. L., Andrich, P., Klimov, P. V., Hassan, J. U., Son, N. T., Janzen, E., Ohshima, T. & Awschalom, D. D. Isolated electron spins in silicon carbide with millisecond coherence times. *Nat. Mater.* **14,** 160–163 (2014).
6. Klauder, J. R. & Anderson, P. W. Spectral Diffusion Decay in Spin Resonance Experiments. *Phys. Rev.* **125,** 912–932 (1962).
7. Hartman, J. S., Berno, B., Hazendonk, P., Kirby, C. W., Ye, E., Zwanziger, J. & Bain, A. D. NMR Studies of Nitrogen Doping in the 4H Polytype of Silicon Carbide: Site Assignments and Spin−Lattice Relaxation. *J. Phys. Chem. C* **113,** 15024–15036 (2009).
8. Hartman, J. S., Narayanan, A. & Wang, Y. Spin-Lattice Relaxation in the 6H Polytype of Silicon Carbide. *J. Am. Chem. Soc.* **116,** 4019–4027 (1994).
9. Schweiger, A. & Jeschke, G. *Principles of Pulse Electron Paramagnetic Resonance*. (Oxford University Press, 2001).
10. Jarmola, A., Acosta, V. M., Jensen, K., Chemerisov, S. & Budker, D. Temperature- and Magnetic-Field-Dependent Longitudinal Spin Relaxation in Nitrogen-Vacancy Ensembles in Diamond. *Phys. Rev. Lett.* **108,** 197601 (2012).
11. Falk, A. L., Klimov, P. V., Buckley, B. B., Ivády, V., Abrikosov, I. A., Calusine, G., Koehl, W. F., Gali, A. & Awschalom, D. D. Electrically and Mechanically Tunable Electron Spins in Silicon Carbide Color Centers. *Phys. Rev. Lett.* **112,** 187601 (2014).
12. Yang, W. & Liu, R.-B. Quantum many-body theory of qubit decoherence in a finite-size spin bath. *Phys. Rev. B* **78,** 085315 (2008).
13. Witzel, W. M., de Sousa, R. & Das Sarma, S. Quantum theory of spectral-diffusion-induced electron spin decoherence. *Phys. Rev. B* **72,** 161306(R) (2005).
14. Van Oort, E. & Glasbeek, M. Optically detected low field electron spin echo envelope modulations of fluorescent N-V centers in diamond. *Chemical Physics* **143,** 131–140 (1990).
15. Guennebaud, G., Jacob, B. & others. *Eigen v3*. (http://eigen.tuxfamily.org, 2010).
16. Yao, W., Liu, R.-B. & Sham, L. J. Theory of electron spin decoherence by interacting nuclear spins in a quantum dot. *Phys. Rev. B* **74,** 195301 (2006).
17. Mims, W. B. Envelope Modulation in Spin-Echo Experiments. *Phys. Rev. B* **5,** 2409–2419 (1972).
18. Maze, J. R., Taylor, J. M. & Lukin, M. D. Electron spin decoherence of single nitrogen-vacancy defects in diamond. *Phys. Rev. B* **78,** 094303 (2008).
19. Zhao, N., Ho, S.-W. & Liu, R.-B. Decoherence and dynamical decoupling control of nitrogen vacancy center electron spins in nuclear spin baths. *Phys. Rev. B* **85**, 115303 (2012).
20. Slichter, C. P. *Principles of Magnetic Resonance*. (Springer Berlin Heidelberg, 1996).